\documentclass[11pt,a4paper]{article}
\usepackage{jheppub}
\usepackage[utf8]{inputenc}
\usepackage{graphicx,fancybox,float,comment,bm}

\usepackage[dvipsnames]{xcolor}
\usepackage{amsmath,amssymb,amsfonts}
\usepackage[bbgreekl]{mathbbol}
\usepackage{multirow}
\usepackage{ulem}

\newcommand{\eps}{\varepsilon}%
\newcommand{\tud}[3]{{#1}{}^{#2}{}_{#3}}%
\newcommand{\tdu}[3]{{#1}{}_{#2}{}^{#3}}%

\title{Spin density from second order transport in  fluids }
\author[1]{Domingo Gallegos, }
\author[2]{Roi Klein, }
\author[2]{Amos Yarom}
\date{September 2022}

\affiliation[1]{Facultad de Ciencias, Universidad Nacional Autonoma de Mexico, Investigacion Cientifica, 04510 Ciudad de Mexico, Mexico}
\affiliation[2]{Department of Physics, Technion, Haifa 32000, Israel}

\abstract{We study the hydrodynamics of a charged fluid with spin degrees of freedom. We argue that, in a conformal field theory in $3+1$ dimensions, the spin density is related to a linear combination of standard second-order transport coefficients, and an additional term quadratic in the gauge- and mixed gauge-gravitational anomaly coefficients. We compute the spin density in several free and holographic theories and compare the results.}

\emailAdd{d.gallegos@ciencias.unam.mx}
\emailAdd{roi.klein@campus.technion.ac.il}
\emailAdd{amos.yarom@gmail.com}

\begin{document}
\maketitle
\section{Introduction}
Relativistic hydrodynamics has played a pivotal role in our understanding of the behavior of the quark gluon plasma \cite{BuszaRajagopalSchee2018,FlorkowskiHellerSpalinski2018,HeinzSchenke2024}. The simultaneous demand for quantitative phenomenological predictions (e.g., \cite{RomatschkeRomatschke2007,SchenkeJeonGale2010}) and a consistent theoretical framework \cite{Muller1967,IsraelStewart1979,BaierRomatschkeSonStarinetsStephanov2008,BemficaDisconziNoronhaKovtun2019} has driven important developments in relativistic-fluid dynamics \cite{PolicastroSonStarinets2001,Kharzeev2006,BhattacharyyaHubenyMinwallaRangamani2008,SonSurowka2009}.

Of particular interest to this work is the recent relation between the relativistic spin current and hydrodynamic behavior of the quark gluon plasma. In \cite{STAR2017,STAR2018,ALICE2020} it was observed that the polarization of $\Lambda$ hyperons is aligned with the angular momentum of the quark gluon plasma. Hydrodynamic descriptions of this behavior rely on a Boltzmann-equation-based analysis 
\cite{Bhadury:2021oat,Weickgenannt:2021cuo,Yi:2021ryh,Das:2022azr,Weickgenannt:2022qvh,Weickgenannt:2022jes,Weickgenannt:2023btk,Weickgenannt:2024ibf,Weickgenannt:2024esg,Daher:2025pfq,Bhadury:2025fil},
partition-function-based analyses
\cite{Becattini:2013fla,Florkowski:2017dyn,Becattini:2018duy,Becattini:2020ngo,Becattini:2021iol,Floerchinger:2021uyo,Buzzegoli:2021wlg,Becattini:2023ouz,Becattini:2025oyi},
or constitutive-relation-based analyses in which the spin current is treated as an additional hydrodynamic current and in which angular momentum conservation supplies the associated conservation law
\cite{Becattini:2007sr,Florkowski:2017ruc,Florkowski:2018fap,Florkowski:2019qdp,Hattori:2019lfp,Fukushima:2020ucl,Li:2020eon,Singh:2020rht,She:2021lhe,Hongo:2021ona,Wang:2021ngp,Gallegos:2021bzp,Singh:2021man,Das:2021aar,Hongo:2022izs,Singh:2022ltu,Gallegos:2022jow,Bhadury:2022ulr,Xie:2023gbo,Valle:2023cqo,Singh:2024rfl,Cartwright:2024dcj,Bhadury:2024whs,Bhadury:2024ckc,Abboud:2025qtg}.
See also
\cite{Voloshin:2004ha,Liang:2004xn,Betz:2007kg,Gallegos:2020otk,Manes:2020zdd,Goncalves:2021ziy,Chen:2021azy,Valle:2021nfv,Dong:2021fxn,Armas:2025fvo,Daher:2025vhl,Armas:2026bmw,Singh:2026wvf,Matthaiakakis:2026qpz,Li:2026vld,Montenegro:2026phf,Palermo:2026mwu}.

As is well known, if the stress tensor is symmetric, angular momentum conservation follows directly from energy-momentum conservation. For this reason, it may seem somewhat puzzling that the angular momentum density is associated with additional dynamical degrees of freedom. More formally, since angular momentum does not commute with translations, we do not expect additional hydrodynamic Goldstone modes associated with the breaking of rotational invariance. In \cite{Li:2020eon,Hongo:2021ona,Hongo:2022izs} this issue was resolved by treating the spin density as slightly massive and computing the associated constitutive relation within the general framework known as ``Hydro $+$' \cite{Stephanov:2017ghc}. Alternatively, in \cite{Gallegos:2021bzp,Gallegos:2022jow,Cartwright:2024dcj}  it was shown that the angular momentum density (and therefore the spin current) is determined by the hydrodynamic variables associated with translation and flavor symmetries. Moreover, treating the spin chemical potential as a dynamical field leads to algebraic constraints that determine it in terms of the other hydrodynamic degrees of freedom. That is, the spin chemical potential is slave to the velocity field, $u^{\mu}$, temperature, $T$, and possible chemical potentials, $\mu$, associated with global (flavor) symmetries.

The spin chemical potential introduced in the previous paragraph induces a spin density. In a conformal fluid, this spin density is characterized by a spin susceptibility $\rho_M$. 
Our main result relates $\rho_M$ to the pressure $P_0$, the gravitational susceptibility (usually denoted by $\kappa$, but denoted here by $\kappa_g$), the nonlinear vortical response coefficient $\lambda_3$ \cite{BaierRomatschkeSonStarinetsStephanov2008,BhattacharyyaHubenyMinwallaRangamani2008}, and the gauge and mixed gauge--gravitational anomaly coefficients $c_A$ and $c_m$:
\begin{equation}
\label{E:rhoMlambda3}
	\rho_M = \frac{\lambda_3}{4} + \frac{\kappa_g}{2} - \frac{\left(4 c_A \mu^3 + 32 \pi^2 c_m T^2\mu\right)^2}{16 P_0}\,.
\end{equation} 
The precise definitions of the various coefficients in this formula will be discussed extensively below. In section \ref{S:findings} we present the constitutive relations for charged spin hydrodynamics with gauge and mixed gauge--gravitational anomalies. Our conventions for the pressure and anomaly coefficients can be found in equations \eqref{E:Pexpansion} and \eqref{E:covariantanom}, respectively. In section \ref{S:main}, we review the nonlinear vortical response and gravitational susceptibility, defined in equation \eqref{E:kappalambda3}. In that section we also derive \eqref{E:rhoMlambda3}. 

Because second-order transport has been studied extensively, equation \eqref{E:rhoMlambda3} allows us to evaluate the spin density in many existing examples.
Indeed, there is a vast literature in which second-order transport coefficients have been computed for a range of free field theories \cite{MooreSohrabi:2010,MooreSohrabi:2012,KovtunShukla:2018,Shukla:2019} and holographic models \cite{Son:2002sd,Policastro:2002se,Kovtun:2005ev,Son:2006em,BhattacharyyaHubenyMinwallaRangamani2008,BaierRomatschkeSonStarinetsStephanov2008,Natsuume:2007ty,Erdmenger:2008rm,Banerjee:2008th,Maeda:2008hn,Hur:2008ju,Matsuo:2009ws,Gynther:2010ed,Amado:2011zx,ArnoldVamanWuXiao:2011,SaremiSohrabi:2011,Landsteiner:2011cp,Landsteiner:2012kd,Megias-Pena:2013,Grozdanov:2014kva,GrozdanovStarinets:2017,Bu:2015ika,Bu:2020teu,GrieningerShukla:2021}.
After developing the formalism, we elaborate on these results in section \ref{S:discussion}. 
We find, for example, that for $SU(N_c)$, $\mathcal{N}=4$ super Yang Mills at infinite 't Hooft coupling and zero chemical potential (for $R$ charge) we have
\begin{equation}
	\rho_M = \frac{N_c^2 T^2}{16}.
\end{equation}
An extension of this result to non vanishing chemical potential, $1/N_c$ corrections or zero coupling can be found in Table \ref{T:various}.
Various detailed computations are relegated to the appendices.

\section{Parity violating hydrodynamics with spin currents}
\label{S:findings}
Hydrodynamics can be thought of as an effective low-energy description of a many-body system at finite temperature. The associated dynamics are captured by conservation laws, e.g., $\nabla_{\mu}T^{\mu\nu}=0$ and $\nabla_{\mu}J^{\mu}=0$ (in the absence of external sources). The dynamical variables are associated with the Goldstone modes of the broken symmetries of the problem, e.g., a velocity field $u^{\mu}$, a temperature field $T$, and a chemical potential $\mu$. There is now a systematic procedure for constructing a hydrodynamic theory from the symmetries of the system \cite{Bhattacharyya:2007vjd,BaierRomatschkeSonStarinetsStephanov2008,Jensen:2012jh,Banerjee:2012iz}.

Angular momentum conservation (in the absence of sources) is encoded in the spin current equation
\begin{equation}
	\nabla_{\mu}S^{\mu\nu\rho} = 2 T^{[\nu\rho]}
\end{equation}
with 
\begin{equation}
	A_{[\mu\nu]} = \frac{1}{2} \left(A_{\mu\nu}-A_{\nu\mu}\right)\,,
\end{equation}
(and, later, circular parenthesis will correspond to symmetrization). Angular momentum conservation supplies an antisymmetric tensor's worth of equations. Therefore, its associated constitutive relations require an auxiliary antisymmetric spin-chemical-potential tensor as a dynamical variable,
\begin{equation}
\label{E:spinCdecomposition}
	\mu^{\alpha\beta} = u^{\alpha}m^{\beta}-u^{\beta}m^{\alpha} + M^{\alpha\beta}\,,
\end{equation}
with $\mu^{\alpha\beta} = -\mu^{\beta\alpha}$ and $m^{\alpha}u_{\alpha} = M^{\alpha\beta}u_{\beta}=0$.
In \cite{Gallegos:2021bzp,Gallegos:2022jow}, the constitutive relations for the spin current in $d+1$ dimensions were derived explicitly using the techniques described above. Not surprisingly, the spin chemical potential is determined algebraically by the ordinary hydrodynamic fields.

In what follows, we build on the results of \cite{Gallegos:2022jow} and extend them to include parity-violating transport in $3+1$ dimensions.
We find
that in a conformal field theory and in the absence of external electromagnetic fields, 
to first order in derivatives,
\begin{align}
\begin{split}
\label{E:fullconstitutive}
	T^{\mu\nu}=& T_{id}^{\mu\nu} + T_A^{\mu\nu} - \sigma_{\mathcal{E}} \mathcal{E}^{[\mu} u^{\nu]} - \sigma_{\tilde{\mathcal{E}}} \tilde{\mathcal{E}}^{\mu \nu}  +\sigma_{{m}} \hat{m}^{[\mu}u^{\nu]} - \frac{1}{2} \sigma_{\tilde{M}} \tilde{\hat{M}}^{[\mu}u^{\nu]} + \sigma_M \hat{M}^{\mu\nu} + \sigma_{\tilde{m}} \tilde{\hat{m}}^{\mu\nu} \\
	&-\eta \sigma^{\mu\nu} + B^{\mu\nu}[\Phi]
	\\
	J^{\mu}  =&  J_{id}^{\mu} + J_A^{\mu} +\sigma_E \mathcal{E}^{\mu} \\
	S^{\lambda\mu\nu} & = S^{\lambda\mu\nu}_{id} + S_d^{\lambda\mu\nu}+ \Phi^{\lambda\mu\nu} 
\end{split}
\end{align}
with the following definitions. 

Expressions with an $id$ subscript correspond to what we refer to as ideal terms:
\begin{align}
\begin{split}
\label{E:idstresscurrent}
        T_{id}^{\mu\nu}&=\eps u^{\mu}u^{\nu}+PP^{\mu\nu}+u^{\mu}P^{\nu}\,, \\
    J_{id}^{\mu}&=\rho u^{\mu}\,,\\
    S_{id}^{\lambda\mu\nu} & = -4 \rho_M u^{\lambda}M^{\mu\nu}
\end{split}
\end{align}
with
\begin{equation}
	P^{\mu\nu} = \eta^{\mu\nu} + u^{\mu}u^{\nu}\,,
\end{equation}
where
\begin{subequations}
\label{E:idealexpansion}
\begin{align}
\label{E:Pexpansion}
	P &= P_0 - \rho_M M_{\alpha\beta}M^{\alpha\beta} \\
\label{E:Pmuexpansion}
	P^{\mu} &=  - 4 \rho_M m_{\alpha}M^{\alpha\mu} \\
\label{E:epsexpansion}
	\epsilon &= \epsilon_{0} - \left( \rho_M + T \frac{\partial \rho_M}{\partial T} + \mu \frac{\partial \rho_M}{\partial \mu} \right) M_{\alpha\beta}M^{\alpha\beta} \,,\\
	\rho& = \frac{\partial P_0}{\partial \mu} - \frac{\partial \rho_M}{\partial \mu} M_{\alpha\beta}M^{\alpha\beta}\,,
\end{align}
\end{subequations}
such that 
\begin{equation}
	\epsilon_0 + P_0 =T \frac{\partial P_0}{\partial T} + \mu \frac{\partial P_0}{\partial \mu}
\end{equation}	
The quantities $\epsilon_0$, $P_0$, and $\rho_M$ are functions of the temperature $T$ and chemical potential $\mu$.
Conformal invariance dictates that $\epsilon_0 = 3 P_0$.

Expressions with an $A$ index are associated with anomalies,
\begin{align}
\begin{split}
\label{E:chiralanom}
    T_A^{\mu\nu}&=\left( -4\mu^3c_{A} -32 \pi^2 c_m \mu T^2 \right) u^{(\mu}\tilde{\Omega}^{\nu)}\,,\\
       J_A^\mu&=\left( - 3\mu^2 c_{A} - 8\pi^2 c_m T^2\right)\tilde{\Omega}^{\mu}\,,
\end{split}
\end{align}
where we use
\begin{equation}
\label{E:defOmega}
	\Omega_{\mu\nu} = P_{\mu}{}^{\rho} P_{\nu}{}^{\sigma}\nabla_{[\rho}u_{\sigma]}
\end{equation}
and, in general, we use the convention
\begin{align}
\begin{split}
\label{E:hodge}
	\tilde{X}^{\alpha} &= \epsilon^{\alpha\beta\gamma\delta}u_{\beta}X_{\gamma\delta} \\
	\tilde{X}^{\alpha\beta} &= \epsilon^{\alpha\beta\gamma\delta}u_{\gamma}X_{\delta}\,.
\end{split}
\end{align}
Note that $\tilde{\tilde{X}}^{\alpha\beta}  = -2 X^{\alpha\beta}$ and $\tilde{\tilde{X}}^{\alpha} = -2 X^{\alpha}$.

The coefficients $c_A$ and $c_m$ denote the strengths of the $U(1)$ and mixed gauge--gravitational anomalies, respectively. In the presence of an external $U(1)$ field strength, $F_{\mu\nu}$, and Riemann curvature $R_{\alpha\beta\gamma\delta}$, we have 
\begin{equation}
\label{E:covariantanom}
	\nabla_{\mu}J^{\mu}=\frac{1}{4}\epsilon^{\mu\nu\rho\sigma}\left(3 c_A F_{\mu\nu}F_{\rho\sigma} + c_m R^{\alpha}{}_{\beta\mu\nu}R^{\beta}{}_{\alpha\rho\sigma}\right)\,.
\end{equation}
Here $J^{\mu}$ is the covariant current rather than the consistent current. (Recall that the consistent current is obtained by directly varying the generating functional. The covariant current is obtained by adding a Bardeen--Zumino improvement term to the consistent current. See, e.g., \cite{Jensen:2012kj} for a discussion.) The expressions \eqref{E:chiralanom} represent the covariant currents as well.

The $\sigma$-type terms in the antisymmetric component of the stress tensor are of the form
\begin{align}
\begin{split}
\label{E:neqterms}
	\mathcal{E}^{\mu} &= -T P^{\mu\alpha}\nabla_{\alpha} \frac{\mu}{T} \\
	\hat{m}^{\mu} &= m^{\mu} - a^{\mu} \\
	\hat{M}^{\mu\nu} &= M^{\mu\nu} + \Omega^{\mu\nu}\,,
\end{split}
\end{align}
Here $a^{\mu}=u^{\alpha}\nabla_{\alpha}u^{\mu}$, $\Omega^{\mu\nu}$ is defined in \eqref{E:defOmega}, and the coefficients $\sigma_X$ are functions of $T$ and $\mu$.
The conductivity is positive, $\sigma_E \geq 0$.
The shear tensor is  given by
\begin{equation}
	\sigma_{\mu\nu} = 2 \nabla_{\langle \mu}u_{\nu\rangle}
\end{equation}
where we have used
\begin{equation}
\label{E:tb}
	A_{\langle \mu\nu \rangle} = \frac{1}{2} P_{\mu}{}^{\alpha} P_{\nu}{}^{\beta} \left(A_{\alpha\beta}+A_{\beta\alpha}\right) - \frac{1}{3} P_{\mu\nu} A_{\alpha}{}^{\alpha}
\end{equation}
and $\eta(T,\mu) \geq 0$ is the shear viscosity. 
The analogous terms in the spin current are encoded in $S_d^{\lambda\mu\nu}$:
\begin{multline}
\label{E:Sd}
	S_{d}^{\lambda\mu\nu}=
	2\sigma_1\sigma^{\lambda[\mu}u^{\nu]}
	+2\widetilde \sigma_1\epsilon^{\mu\nu}{}_{\rho\sigma}\sigma^{\lambda\rho}u^\sigma
	-2\sigma_2g^{\lambda[\mu}\mathcal E^{\nu]}
	-2\widetilde \sigma_2u^{[\lambda}\widetilde{\mathcal E}^{\mu\nu]}
	\\
	-2\sigma_3\left(u^\lambda u^{[\mu}\mathcal E^{\nu]}+\frac13g^{\lambda[\mu}\mathcal E^{\nu]}\right)
	-2\widetilde \sigma_3\left(u^\lambda\widetilde{\mathcal E}^{\mu\nu}-u^{[\lambda}\widetilde{\mathcal E}^{\mu\nu]}\right)\,,
\end{multline}
where the $\sigma_i$ and $\widetilde{\sigma}_i$ are functions of $T$ and $\mu$.

The tensor $\Phi^{\lambda\mu\nu}$ encodes the following terms,
\begin{align}
\begin{split}
\label{E:Phiterm}
	\Phi^{\lambda\mu\nu}=&
	2\zeta^{\left(1\right)}g^{\lambda[\mu}u^{\nu]}
	+\frac{1}{3}\tilde{\zeta}^{\left(1\right)}\eps^{\lambda\mu\nu\rho}u_{\rho}
    	+2\zeta_{1}^{\left(2\right)}M^{[\mu\nu}u^{\lambda]}
	+2\zeta_{2}^{\left(2\right)}\left(M^{\mu\nu}u^{\lambda}-M^{[\mu\nu}u^{\lambda]}\right)
    \\&
    +2\tilde{\zeta}_{1}^{\left(2\right)}g^{\lambda[\mu}\tilde{M}^{\nu]}
    +2\tilde{\zeta}_{2}^{\left(2\right)}\left(u^{\lambda}u^{[\mu}\tilde{M}^{\nu]}
    +\frac{1}{3}g^{\lambda[\mu}\tilde{M}^{\nu]}\right) 
	+2\gamma_1\sigma^{\lambda[\mu}u^{\nu]}
	+2\widetilde\gamma_1\epsilon^{\mu\nu}{}_{\rho\sigma}\sigma^{\lambda\rho}u^\sigma
	\\&
	-2\gamma_2g^{\lambda[\mu}\mathcal E^{\nu]}
	-2\widetilde\gamma_2u^{[\lambda}\widetilde{\mathcal E}^{\mu\nu]}
	-2\gamma_3\left(u^\lambda u^{[\mu}\mathcal E^{\nu]}+\frac13g^{\lambda[\mu}\mathcal E^{\nu]}\right)
	-2\widetilde\gamma_3\left(u^\lambda\widetilde{\mathcal E}^{\mu\nu}-u^{[\lambda}\widetilde{\mathcal E}^{\mu\nu]}\right).
\end{split}
\end{align}
Here the $\zeta_i^{(j)}$, $\sigma_i$, $\gamma_i$, and their tilded counterparts depend on $T$ and $\mu$. The partner tensor $B^{\mu\nu}[\Phi]$ reads
\begin{equation}
	B^{\mu\nu}[\Phi] = \frac{1}{2} \nabla_{\lambda}\left(\Phi^{\lambda\mu\nu} - \Phi^{\mu\lambda\nu} - \Phi^{\nu\lambda\mu}\right)\,.
\end{equation}
The structure of $B^{\mu\nu}$ ensures that $\Phi^{\mu\nu\rho}$ will not contribute to the equations of motion.

We have referred to the constitutive relations \eqref{E:fullconstitutive} as first order in derivatives even though we have determined the antisymmetric components of the stress tensor to second order in derivatives. The reason for this is that the equations of motion relate the derivative of the spin current to the antisymmetric component of the stress tensor. This algebraic relation requires the second-order constitutive relations for the antisymmetric stress tensor in order to obtain the equations of motion through second derivative order.

A derivation of these constitutive relations, including an electric field and nonconformal contributions, is given in Appendix \ref{A:lightningreview}. It closely follows the methods discussed in \cite{Gallegos:2021bzp,Gallegos:2022jow}. In the remaining subsections, we briefly discuss the role of the various terms in the constitutive relations.

\subsection{Ideal hydrodynamics}
The ideal terms $T_{id}^{\mu\nu}$, $J_{id}^{\mu}$ and $S_{id}^{\mu\nu\rho}$ are referred to as ideal terms since they do not include explicit derivatives of the hydrodynamic variables. As discussed in \cite{Gallegos:2022jow}, in equilibrium, one may define the entropy density, $s$, spin charge density, $\rho^{\alpha\beta}$, and charge density, $\rho$, as derivatives of a pressure function $P(T,\mu,\mu^{\alpha\beta},u^{\alpha})$,
\begin{equation}
\label{E:firstlaw}
	dP = s dT + \frac{1}{2} \rho_{\alpha\beta}d\mu^{\alpha\beta} + \rho d\mu + P_{\mu}du^{\mu}\,.
\end{equation}
The energy density may be defined via the Gibbs Duhem relation
\begin{equation}
\label{E:GibbsDuhem}
	\epsilon  = sT+\mu\rho+\frac{1}{2} \mu^{\alpha\beta}\rho_{\alpha\beta} -P \,.
\end{equation}
With these definitions, the stress tensor and current will take the form in \eqref{E:idstresscurrent} and the spin current will take the form
\begin{equation}
\label{E:Sideal}
	S_{id}^{\mu\nu\rho} = u^{\mu} \rho^{\nu\rho}\,.
\end{equation}

To obtain the expressions in \eqref{E:idealexpansion}, we treat the spin chemical potential as a first order in derivative object \cite{Gallegos:2021bzp,Gallegos:2022jow} and expand the pressure in a(n implicit) derivative expansion. The leading term, denoted by $P_0$, represents the pressure in the absence of a spin chemical potential. At second derivative order, conformal invariance permits a single contribution to the pressure involving the spin chemical potential. Its coefficient, $\rho_M(T,\mu)$, also controls the spin density.
 (Without conformal invariance, additional terms would appear on the right-hand side of that equation.)
A straightforward computation now gives 
\begin{align}
\begin{split}
\label{E:idealcharges}
	s &= s_0 - \frac{\partial \rho_M}{\partial T} M_{\alpha\beta}M^{\alpha\beta}\,, \\
	\rho &= \rho_0 - \frac{\partial \rho_M}{\partial\mu}M_{\alpha\beta}M^{\alpha\beta}\,,\\
	\rho^{\mu\nu} & = -4 \rho_M M^{\mu\nu} \,,
\end{split}
\end{align}
where $s_0 = \frac{\partial P_0}{\partial T}$ and $\rho_0 = \frac{\partial P_0}{\partial \mu}$. Note that $\rho_M$ determines the spin density.

\subsection{The chiral and mixed gauge-gravitational anomaly}
The pioneering works \cite{Erdmenger:2008rm,Banerjee:2008th,SonSurowka2009} showed that the chiral anomaly leads to macroscopic phenomena such as the chiral vortical and chiral magnetic effects \cite{Kharzeev2006}. The former is encoded in the anomaly-type terms in \eqref{E:chiralanom}. As it turns out, the chiral anomaly does not play a special role in the conservation laws for angular momentum.

Apart from the chiral anomaly, the theory may also possess a mixed axial-gravitational anomaly. It, too, can contribute to transport \cite{Landsteiner:2011iq,Landsteiner:2011cp,Jensen:2012kj,Golkar:2012kb,Jensen:2013kka,Jensen:2013rga} and have a macroscopically observable effect \cite{Gooth:2017mbd}. The mixed anomaly is often described as an anomaly associated with chiral and diffeomorphism symmetries. Somewhat more precisely, it mixes chiral, diffeomorphism, and local Lorentz symmetries \cite{Jensen:2013kka}. It would be interesting to understand its  effect on theories with background torsion. We leave such studies to future work.\footnote{Putative torsion anomalies as encoded in Nieh--Yan terms \cite{NiehYan1982,Chandia:1997hu} can be removed by a local counterterm \cite{Kreimer:1999,NiehYanLandsteiner,ErdmengerNiehYan} which makes them cohomologically trivial and therefore not genuine anomalies. Retaining such terms corresponds to an explicit breaking of the symmetry, or equivalently to a non-invariant choice of local counterterm. See \cite{Peeters:1999,ErdmengerNiehYan,HoyosNiehYan} for a study of Nieh--Yan terms on manifolds with a boundary.}

\subsection{Hydrodynamic improvement terms (``pseudogauges'')}
\label{SS:pseudogauge}
The stress tensor may be improved. Consider the stress tensor $T^{\mu\nu}$ and spin current $S^{\mu\nu\rho}$. In Minkowski space and in the absence of external electromagnetic fields, the conservation equations (Ward identities) are
\begin{equation}
\label{E:Conservationfirst}
	\nabla_{\mu}T^{\mu\nu} = 0\,,
	\qquad
	\nabla_{\mu}S^{\mu\nu\rho} = 2 T^{[\nu\rho]} \,.
\end{equation}
As it turns out, we can always define an improved stress tensor $T^{\prime\mu\nu}$ and an improved spin current $S^{\prime\mu\nu\rho}$ such that
\begin{align}
\begin{split}
\label{E:pseudogauge}
	T^{\prime\,\mu\nu} &= T^{\mu\nu} + B^{\mu\nu}[J] \\ 
	S^{\prime\,\alpha\mu\nu} &= S^{\alpha\mu\nu} + J^{\alpha\mu\nu} 
\end{split}
\end{align}
which will satisfy
\begin{equation}
\label{E:Conservation}
	\nabla_{\mu}T^{\prime\,\mu\nu} = 0\,,
	\qquad
	\nabla_{\mu}S^{\prime\,\mu\nu\rho} = 2 T^{\prime[\nu\rho]} \,,
\end{equation}
for any $J^{\alpha\beta\gamma}$.

In quantum field theory, it is common to construct $J^{\alpha\beta\gamma}$ from the fundamental fields and their derivatives. In hydrodynamics, it may instead be constructed from the hydrodynamic fields.
Note that in \eqref{E:fullconstitutive} the $\Phi^{\mu\nu\rho}$ terms defined in \eqref{E:Phiterm} come with a partner contribution to the stress tensor, $B^{\mu\nu}(\Phi)$. The correlated pair $(B[\Phi],\Phi)$ cancels under $J^{\mu\nu\rho} = -\Phi^{\mu\nu\rho}$. We will refer to such pairs as pure improvement terms.

Note that it is always possible to choose $J^{\alpha\mu\nu}=-S^{\alpha\mu\nu}$. In this case, the spin current will vanish and its contributions to the equations of motion will be rotated into the stress tensor. If we denote the stress tensor and spin current obtained from this improvement by $T_C^{\mu\nu}$ and $S_C^{\mu\nu\rho}$, respectively, then
\begin{align}
\begin{split}
\label{E:symmetricconstitutive}
	T_C^{\mu\nu}=& T_{id}^{\mu\nu} + T_A^{\mu\nu} - \sigma_{\mathcal{E}} \mathcal{E}^{[\mu} u^{\nu]} - \sigma_{\tilde{\mathcal{E}}} \tilde{\mathcal{E}}^{\mu \nu}  +\sigma_{{m}} \hat{m}^{[\mu}u^{\nu]} - \frac{1}{2} \sigma_{\tilde{M}} \tilde{\hat{M}}^{[\mu}u^{\nu]} + \sigma_M \hat{M}^{\mu\nu} + \sigma_{\tilde{m}} \tilde{\hat{m}}^{\mu\nu} \\
	&-\eta \sigma^{\mu\nu} - B^{\mu\nu}[S_{id} +S_d]
	\\
	J^{\mu}  =&  J_{id}^{\mu} + J_A^{\mu} +\sigma_E \mathcal{E}^{\mu} \\
	S_C^{\mu\nu\rho} & = 0\,,
\end{split}
\end{align}
and the equations of motion now read
\begin{equation}
	T_C^{[\mu\nu]}=0
	\qquad
	\nabla_{\mu}T_C^{\mu\nu}=0\,.
\end{equation}
Note that $T_C^{\mu\nu}$ is independent of the improvement term (it is ``pseudo-gauge invariant''). 
We may also define the manifestly symmetric and conserved stress tensor
\begin{equation}
	T_{s}^{\mu\nu} = T_C^{\mu\nu} - T_C^{[\mu\nu]}\,.
\end{equation}

Recall that our explicit expressions for the constitutive relations in \eqref{E:fullconstitutive} were truncated so that the equations of motion are correct through first order in the derivative expansion: the charged current, symmetric stress tensor and spin current were expanded to first order in derivatives, while the antisymmetric component of the stress tensor was expanded to second order in derivatives. Therefore, in the present construction we have
\begin{equation}
	\nabla_{\mu}T_C^{\mu\nu} = \mathcal{O}(\nabla^3)
	\qquad
	T_C^{[\mu\nu]} = \mathcal{O}(\nabla^3)\,.
\end{equation}

Note that for a first order  $J^{\lambda\mu\nu}$, $B^{\mu\nu}[J]$ also contains second order terms which are symmetric in $\mu$ and $\nu$. These terms are not needed to determine the equations of motion through second order, since their divergence begins at third order. Nevertheless, the first order spin-current contribution is not separately observable from the independent second-order coefficients of the improvement term invariant $T_C^{\mu\nu}$. We will make use of this observation in Section \ref{S:main}.

\subsection{Hydrodynamic frames}
\label{SS:frames}
Like any low-energy effective description, hydrodynamics has a field-redefinition ambiguity, commonly described in terms of ``fluid frames.'' For instance, given a velocity field $u^{\mu}$ and temperature $T$, one may alternatively choose new fields $\tilde{T}=T+u^{\alpha}\nabla_{\alpha}T$ and $\tilde{u}^{\mu}=u^{\mu}+u^{\alpha}\nabla_{\alpha}u^{\mu}$, which coincide with $T$ and $u^{\mu}$ in equilibrium but differ away from equilibrium. That is, the pairs $(u^{\mu},T)$ and $(\tilde{u}^{\mu},\tilde{T})$ agree at leading order in the derivative expansion but differ once gradient corrections are included.

Frame transformations do not affect physical results provided that the derivative expansion is treated perturbatively.\footnote{When applying hydrodynamic equations to a particular problem, one often does not enforce a strictly perturbative treatment of the derivative expansion. This has a variety of effects on causality and well-posedness of the initial value problem \cite{Bemfica:2017wps,Bemfica:2019knx,Kovtun:2019hdm}.} The stress tensor and other physical observables are invariant under changes of fluid frame \cite{Bhattacharya:2011eea}.

Although a frame-invariant formulation is sometimes possible, it is often useful to choose a frame suited to the problem at hand. Since the frame ambiguity is tied to the number of dynamical degrees of freedom, fixing the frame requires the same number of constraints. For instance, in hydrodynamics with a symmetric stress tensor, the Landau frame is commonly used. If we decompose the stress tensor into a zeroth-order term and a subleading correction, $T^{\mu\nu}=T^{(0)\mu\nu}+\tau^{\mu\nu}$, then the Landau frame corresponds to a choice of velocity and temperature such that
\begin{equation}
	T^{\mu\nu}u_{\nu} = T^{(0)\mu\nu} u_{\nu}\,.
\end{equation}
These four equations fix the field redefinition ambiguity of the four hydrodynamic degrees of freedom, $u^{\mu}$ and $T$. 

In the presence of a charge current, we have an additional chemical potential as a dynamical variable. As a result, fixing the frame requires an additional condition on the hydrodynamic variables.
Likewise, an antisymmetric spin chemical potential requires six additional conditions in order to remove the associated field redefinition ambiguity. We may fix these ambiguities by placing constraints on the antisymmetric stress tensor. Because the stress tensor enters the equations of motion algebraically, cf.\ \eqref{E:Conservation}, the associated field redefinitions are of second derivative order. Thus, we can shift $\hat{M}^{\mu\nu}$ and $\hat{m}^{\mu}$ by second order in derivative expressions to ensure that the non-improvement second order in derivative antisymmetric stress tensor receives no contributions except for possible pure improvement terms. 

In writing \eqref{E:fullconstitutive}, we have fixed a hybrid frame in which only the dissipative terms are in the Landau frame, while the remaining terms are in the hydrostatic frame. We also allowed pure improvement terms. See \cite{Gallegos:2022jow}. Other constitutive relations in which the second-order antisymmetric stress tensor takes a different form are also possible. When the derivative expansion is treated perturbatively, all such frames yield the same physical results.

\subsection{Angular momentum conservation}
We have chosen a framework in which the stress tensor is not manifestly symmetric. As a result, angular momentum conservation yields an additional set of equations of motion of the form \eqref{E:Conservationfirst}, which in our setup read 
\begin{subequations}
\label{E:asym}
\begin{equation}
\label{E:vectorasym}
	\sigma_m \hat{m}^{\mu} - \frac{1}{2} \sigma_{\tilde{M}} \tilde{\hat{M}}^{\mu}-\sigma_{\mathcal{E}} \mathcal{E}^{\mu} 
	-4 \rho_M M^{\mu\rho}\hat{m}_{\rho} +P^{\mu}{}_{\alpha} u_{\beta} \nabla_{\lambda}S^{\lambda\alpha\beta}_d =  0
\end{equation}
and
\begin{multline}
\label{E:tensorasym}
	\sigma_M \hat{M}^{\mu\nu} + \sigma_{\tilde{m}} \tilde{\hat{m}}^{\mu\nu} - \sigma_{\tilde{\mathcal{E}}}\tilde{\mathcal{E}}^{\mu\nu} 
	+2\rho_{M}P^{\alpha}{}_{\mu} P^{\beta}{}_{\nu} \dot{M}^{\alpha\beta}
	\\
	+2\rho_{M}\left(\nabla_{\gamma}u^{\gamma}\right) M^{\mu\nu}+2\dot{\rho}_M M^{\mu\nu} -\frac{1}{2}P^{\mu}{}_{\alpha}P^{\nu}{}_{\beta} \nabla_{\lambda}S_d^{\lambda\alpha\beta} = 0\,.
\end{multline}
\end{subequations}
Recall from \eqref{E:spinCdecomposition} that $m^{\mu}$ and $M^{\mu\nu}$ are the components of the spin chemical potential parallel and orthogonal to the velocity, respectively. Their hatted versions are combinations of the spin chemical potential and other thermodynamic variables that vanish in equilibrium; see \eqref{E:neqterms}. Tilded quantities are Hodge duals of non-tilded quantities, cf.\ \eqref{E:hodge}, while dots denote local time derivatives:
\begin{equation}
	\dot{X} = u^{\lambda}\nabla_{\lambda}X\,.
\end{equation}
The vector $\mathcal{E}^{\mu}$ is associated with the gradient of the chemical potential and, in the absence of an external electric field, also vanishes in equilibrium; see \eqref{E:neqterms}. The expression for $S_d^{\lambda\mu\nu}$ can be found in \eqref{E:Sd}.

Equations \eqref{E:asym} contain both first order and second order terms. The first three terms on the left hand side of \eqref{E:vectorasym} and the  first three terms on the left hand side of \eqref{E:tensorasym} are first order in derivatives, whereas the remaining terms are second order in derivatives. We stress that the form of the second order terms is frame dependent and that a different frame may add further second order (or higher order) terms to \eqref{E:vectorasym}.  As emphasized earlier, such a frame choice will not affect the expectation values of the stress tensor and charge current as long as the equations of motion are used perturbatively. That is: if we solve the leading order equations of motion first and treat the subleading ones as corrections to the former. If instead we solve \eqref{E:asym} without a perturbative expansion, its validity may be questioned, and additional terms may be needed to describe a particular physical effect; see, e.g., \cite{Bemfica:2017wps,Bemfica:2019knx,Kovtun:2019hdm}. There are of order ten terms that may be added to \eqref{E:asym} at second order in derivatives using a frame transformation, and many more terms that can be added at third order in derivatives.

Let us examine each equation in \eqref{E:asym} more closely. Starting from the vector equation, \eqref{E:vectorasym}, we note that in the absence of a charge current and parity breaking terms, we obtain
\begin{equation}
	\sigma_m \hat{m}^{\mu} = \mathcal{O}(\nabla^2)\,,
\end{equation}
where we have used $M^{\nu\rho}m_{\rho} = \mathcal{O}(\nabla^2)$. Thus, in this setup and to leading order in the derivative expansion, we find that $\hat{m}^{\mu}=0$, implying that
\begin{equation}
	\mu^{\mu\beta}u_{\beta} = a^{\mu}\,.
\end{equation}
The component of the spin chemical potential along the velocity equals the acceleration, as it does in hydrostatic equilibrium. In the presence of a parity preserving conserved current, we find that the spin chemical potential deviates from its equilibrium value,
\begin{equation}
	\sigma_m  \mu^{\mu\beta}u_{\beta}  = \sigma_m a^{\mu} - \sigma_{\mathcal{E}} T P^{\mu\beta} \nabla_{\beta} \frac{\mu}{T}\,.
\end{equation}

The tensor equation behaves similarly. In the absence of parity-breaking terms and a conserved current, we find that the spin chemical potential takes its equilibrium value:
\begin{equation}
	P^{\mu\alpha}P^{\nu\beta}\mu_{\alpha\beta} = -\Omega^{\mu\nu} + \mathcal{O}(\nabla^2)\,.
\end{equation}
As was the case for the vector component of the antisymmetric equation, this equation gets modified in the presence of charge, and couples to the vector components once parity breaking terms are allowed.  For example, in the absence of charge and to leading order in derivatives, we have
\begin{align}
\begin{split}
	\sigma_m \hat{m}^{\mu} - \frac{1}{2} \sigma_{\tilde{M}} \epsilon^{\mu\nu\alpha\beta}u_{\nu}\hat{M}_{\alpha\beta} &= \mathcal{O}(\nabla^2) \\
	\sigma_M \hat{M}^{\mu\nu} + \sigma_{\tilde{m}}\epsilon^{\mu\nu\rho\sigma}u_{\rho}\hat{m}_{\sigma} &= \mathcal{O}(\nabla^2)\,.
\end{split}
\end{align}
Substituting either equation into the other, we find $\hat{m}^{\mu}=0$ and $\hat{M}^{\mu\nu}=0$, provided that none of the coefficients $\sigma_X$ is perturbatively small; see \cite{Gallegos:2020otk}. Thus, the spin chemical potential equals the thermal vorticity even away from thermal equilibrium. This result is modified in the presence of a $U(1)$ chemical potential in a manner similar to the parity preserving case.

Finally, in light of \cite{Wagner:2024fhf}, let us consider the second order terms, even though they're not unique. In the absence of parity breaking terms and a $U(1)$ chemical potential, we find, in the tensor sector:
\begin{equation}
\label{E:2ndorder}
	\dot M^{\mu\nu} +2\left(\frac{\dot\rho_M}{\rho_M}+ \nabla_{\gamma}u^{\gamma}\right)M^{\mu\nu} = - \frac{\sigma_M}{2\rho_M}\left(M^{\mu\nu}+\Omega^{\mu\nu}\right) -\frac{\sigma_1}{2\rho_M}\sigma^{\lambda[\mu}\Omega_{\lambda}{}^{\nu]}\,.
\end{equation}
Thus $M^{\mu\nu}$ relaxes toward $-\Omega^{\mu\nu}$, while the expansion of the fluid, the variation of the susceptibility, and the shear and vorticity provide additional corrections. To make contact with \cite{Wagner:2024fhf}, consider fluctuations of the spin chemical potential in a  background fluid with constant velocity, temperature, and chemical potential. In this limit \eqref{E:asym} reduces to
\begin{equation}
\label{E:spindecay}
 	\dot M^{\mu\nu} = -\frac{\sigma_M}{2\rho_M} M^{\mu\nu}\,.
\end{equation}
Accordingly, the tensor component of the spin chemical potential decays on the characteristic timescale
\begin{equation}
	\tau_s=\frac{2\rho_M}{\sigma_M}\,,
\end{equation}
(assuming it is positive). A similar linearized analysis gives $m^{\mu}=0$ for the vector equation. Equations of this type are used in \cite{Wagner:2024fhf}, where the corresponding relaxation coefficients are evaluated microscopically in kinetic theory. Equation \eqref{E:2ndorder} is an extension of this relaxation equation to a dynamical fluid background.

As mentioned earlier, when solving the equations of motion \eqref{E:2ndorder} perturbatively, the expectation value of the antisymmetric part of the energy momentum tensor will not depend on the second order terms in \eqref{E:2ndorder}. Here, ``perturbative'' means that the subleading equations of motion provide corrections to the leading-order solution. So, for example, we would need to expand the velocity field, $u^{\mu} = u^{\mu}_{(0)}+ u^{\mu}_{(1)} + \ldots$ where $u^{\mu}_{(0)}$ is the solution to the leading order in derivative (inviscid) equations of motion and $u_{(1)}^{\mu}$ is a solution to the leading viscous terms.

Because the hydrodynamic derivative expansion is perturbative, the leading-order equations of motion may be used to choose a basis for the subleading terms. For instance, the leading-order equations of motion read
\begin{align}
\begin{split}
\label{E:zeroEOM}
	u^{\lambda}\nabla_{\lambda}T &= - T \frac{\partial P}{\partial \epsilon} \nabla_{\alpha}u^{\alpha} \\
	u^{\lambda}\nabla_{\lambda}\frac{\mu}{T} & = - \frac{1}{T} \frac{\partial P}{\partial \rho} \nabla_{\alpha}u^{\alpha} \\ 	
	u^{\lambda}\nabla_{\lambda}u^{\mu} &= \frac{\rho}{\epsilon+P}  \mathcal{E}^{\mu} - \frac{1}{T} P^{\mu\nu} \nabla_{\nu}T\,,
\end{split}
\end{align}
we may replace all instances of $\dot{T}$, $\dot{\mu}$ and $a^{\mu}$ with the right-hand sides of \eqref{E:zeroEOM}. Similarly, we can use the leading order equations of motion in \eqref{E:asym} to obtain $M^{\mu\nu}+\Omega^{\mu\nu}=\alpha_M \tilde{\mathcal{E}}^{\mu\nu}$, where 
\begin{equation}
\label{E:alphaM}
	\alpha_M = \frac{\sigma_m \sigma_{\tilde{\mathcal{E}}} - \sigma_{\tilde{m}}\sigma_{\mathcal{E}}}{\sigma_m \sigma_M - \sigma_{\tilde{M}}\sigma_{\tilde{m}}}
\end{equation}
is a rational function of the various $\sigma_X$. We can now use this equality to replace $\dot{M}^{\mu\nu}$ on the right hand side of \eqref{E:asym} with 
\begin{equation}
\label{E:removedott}
	\dot{M}^{\mu\nu} = u^{\lambda}\nabla_{\lambda} \left(\alpha_M \tilde{\mathcal{E}}^{\mu\nu}\right) - \dot{\Omega}^{\mu\nu}\,.
\end{equation}

Although hydrodynamics is constructed as a derivative expansion, its equations of motion are often treated as exact, nonperturbative equations. A prime example is \eqref{E:spindecay} where, as we explained in detail, the structure of the derivative expansion is abused. Nevertheless, even if interpreting the truncated derivative expansion as an evolution equation is not systematically justified, the resulting spin-relaxation picture successfully explains the observed data, as demonstrated in \cite{Wagner:2024fhf}.

\subsection{An aside on $\mathcal{N}=1$ superconformal theories}
\label{SS:SUSY}
In \cite{Cartwright:2024dcj} it was shown that there is a natural pure improvement term for $\mathcal{N}=1$ superconformal theories in four dimensions where 
\begin{equation}
	S^{\mu\nu\rho}=\frac{1}{2} \epsilon^{\mu\nu\rho\lambda} J_{\lambda}\,,
\end{equation}
and $J^{\mu}$ is the $R$ current.
This set of constitutive relations can be obtained from \eqref{E:fullconstitutive} by choosing the improvement terms such that
\begin{align}
\begin{split}
    	\tilde \zeta^{(1)} &= \frac{3}{2} \rho_0 \, , \qquad  
	\zeta^{(1)} = 0  \, , 
	\\
        \tilde \zeta^{(2)}_1 &= 0  \, , \qquad 
        \zeta^{(2)}_1 = \frac{3}{2} \left(3 \mu^2 c_A + 8 \pi^2 c_m T^2 \right) \, , 
        \\
        \tilde \zeta^{(2)}_2 &=0  \, , \qquad  
         \zeta^{(2)}_2 = 0\, ,
         \\
         \tilde{\gamma}_1 & = 0 \,,\qquad
         \gamma_1=0 \,,
         \\
         \tilde{\gamma}_2 &= \frac{3}{4}\sigma_E +  \frac{3}{2} \left(3 \mu^2 c_A + 8 \pi^2 c_m T^2 \right) \alpha_M \,,\qquad
         \gamma_2=0 \,,
         \\
         \tilde{\gamma}_3 & = 0 \,, \qquad
         \gamma_3=0\,.
\end{split}
\end{align}
(where $\alpha_M$ was defined in \eqref{E:alphaM})
which leads to 
\begin{equation}
	\Phi^{\mu\nu\rho} = \frac{1}{2} \epsilon^{\mu\nu\rho}{}_{\lambda}J^{\lambda} \equiv \tilde{J}^{\mu\nu\rho}\,.
\end{equation}
To obtain the supersymmetric spin current, we carry out an additional transformation characterized by
\begin{equation}
	J_{SUSY}^{\mu\nu\rho} = -S_{id}^{\mu\nu\rho} - S_d^{\mu\nu\rho}\,.
\end{equation}
As a result, the stress tensor will be shifted by $B[J_{SUSY}]$ and the spin current will be shifted by $J^{\mu\nu\rho}_{SUSY}$. The resulting constitutive relations will now read
\begin{align}
\begin{split}
\label{E:susyconstitutive}
	T^{\mu\nu}=& T_{id}^{\mu\nu} + T_A^{\mu\nu} - \sigma_{\mathcal{E}} \mathcal{E}^{[\mu} u^{\nu]} - \sigma_{\tilde{\mathcal{E}}} \tilde{\mathcal{E}}^{\mu \nu}  +\sigma_{{m}} \hat{m}^{[\mu}u^{\nu]} - \frac{1}{2} \sigma_{\tilde{M}} \tilde{\hat{M}}^{[\mu}u^{\nu]} + \sigma_M \hat{M}^{\mu\nu} + \sigma_{\tilde{m}} \tilde{\hat{m}}^{\mu\nu} \\
	&-\eta \sigma^{\mu\nu} + B^{\mu\nu}[ \tilde{J} +J_{SUSY}]  
	\\
	J^{\mu}  =&  J_{id}^{\mu} + J_A^{\mu} +\sigma_E \mathcal{E}^{\mu} \\
	S^{\mu\nu\rho} & = \frac{1}{2} \epsilon^{\mu\nu\rho\lambda} J_{\lambda}
\end{split}
\end{align}
Since the $R$ current contains the axial fermion current, this identification may be viewed as a supersymmetric completion of the familiar relation between the spin current and the Hodge dual of the axial current.

\section{Spin density from second order transport}
\label{S:main}

The constitutive relations in \eqref{E:fullconstitutive} describe the fluid equations of motion through second derivative order. One may, of course, include higher order corrections to the equations of motion, controlled by numerous additional transport coefficients; see, e.g., \cite{Bhattacharyya:2007vjd,Erdmenger:2008rm,Banerjee:2008th,Banerjee:2012iz}:
\begin{multline}
\label{E:kappalambda3}
	T_s^{\mu\nu} = \ldots 
		+\lambda_1 \sigma^{\alpha \langle\mu}\sigma_{\alpha}{}^{\nu\rangle}
		+\lambda_2 \sigma^{\alpha\langle\mu}\Omega_{\alpha}{}^{\nu\rangle}
		\\
		+\lambda_3 \Omega^{\alpha \langle\mu}\Omega_{\alpha}{}^{\nu\rangle}
		 + \kappa_g \left(R^{\langle \mu\nu\rangle } - 2  u_{\alpha}R^{\alpha\langle \mu\nu \rangle\beta}u_{\beta} \right)
		+\ldots + \mathcal{O}(\nabla^3)\,.
\end{multline}
Here, the first $\ldots$ denote zeroth- and first order terms in the derivative expansion, while the second $\ldots$ denote additional two derivative tensor structures, which we omit for brevity.
We denote the Ricci and Riemann tensors by $R_{\mu\nu}$ and $R_{\mu\nu\rho\sigma}$, respectively. These second order terms will contribute to the equations of motion at third order in derivatives.

The main result of this section relates the spin susceptibility in conformal field theories to $\lambda_3$, $\kappa_g$, and the anomaly coefficients through \eqref{E:rhoMlambda3}, reproduced here for convenience:
\begin{equation*}
\tag{\ref{E:rhoMlambda3}}
	\rho_M = \frac{\lambda_3}{4} + \frac{\kappa_g}{2} - \frac{\left(4 c_A \mu^3 + 32 \pi^2 c_m T^2\mu\right)^2}{16 P_0}\,.
\end{equation*} 
We derive \eqref{E:rhoMlambda3} by considering a configuration in hydrostatic equilibrium, for which only a subset of the second order terms contributes to the stress tensor. We then identify the contribution of $\rho_M$ to the constitutive relations with $\lambda_3$, $\kappa_g$, and the anomaly coefficients.

Hydrostatics concerns fluid configurations that are in equilibrium under time independent external forces. In the absence of contorsion, these forces are encoded by an electromagnetic potential $A_{\mu}$ and the metric $g_{\mu\nu}$. The time independence of the configuration is characterized by a timelike Killing vector $V^{\mu}$ and a compensating gauge parameter $\Lambda_{V}$ 
such that
\begin{equation}
\label{E:timeindependence}
    \pounds_V A_\mu+\partial_\mu\Lambda_V=0\,,
    \qquad
    \pounds_V e^a{}_\mu-\theta_V{}^a{}_b e^b{}_\mu=0\,.
\end{equation}
In such a background, the hydrostatic values of the hydrodynamic variables are
\begin{equation}
\label{E:uTtoK}
    \frac{u^\mu}{T}=V^\mu\,,
    \qquad
    \frac{\mu}{T}= V^\mu A_\mu+\Lambda_{V}\,.
\end{equation}
As a result, the fluid shear and expansion vanish:
\begin{equation}
\label{E:hconstraints}
   \nabla_\mu u^\mu=0\,,
    \qquad
    \sigma^{\mu\nu}=0\,.
\end{equation}
We emphasize that, although turning on the sources $g_{\mu\nu}$ and $A_{\mu}$ is necessary to define hydrostatic configurations, we are ultimately interested in the implications of hydrostatics for fluid flow in a flat background with vanishing gauge field.

The generating functional $W_0$ for connected current correlators in hydrostatic equilibrium may be constructed explicitly as a derivative expansion \cite{Banerjee:2012iz,Jensen:2012jh}. Following \cite{BhattacharyyaDavidThakur:2013,Bhattacharyya:2014PartitionExample,KovtunShukla:2018}, we find that
\begin{multline}
\label{E:WnoK}
	W_0 = \int P(T,\mu) \sqrt{-g} d^4x + W_A 
	\\
	+ \int \left( \lambda_3^h  \,\Omega^{\alpha\beta}\Omega_{\alpha\beta} -\frac{\kappa_g}{2}  (R+6 a_{\alpha}a^{\alpha} ) +3 \frac{\partial \kappa_g}{\partial \mu} E_{\alpha}a^{\alpha} + \ldots \right) \sqrt{-g} d^4x\,.
\end{multline}
The first term is the pressure term. The second term, $W_A$, reproduces anomalous transport and leads to the hydrostatic constitutive relations in \eqref{E:chiralanom} once appropriate Bardeen-Zumino improvement terms are added \cite{Jensen:2012kj,Jensen:2013kka,Jensen:2013rga}. The explicit expressions in the second line are the only ones that will contribute to the transverse symmetric stress tensor in the hydrostatic frame. Other second order terms are denoted by $\ldots$. Here $E^{\mu}=F^{\mu\nu}u_{\nu}$ denotes the local electric field and $R$ the Ricci scalar.
Using $\langle T^{\mu\nu} \rangle = \frac{2}{\sqrt{-g}} \frac{\delta W_0}{\delta g^{\mu\nu}}$ together with \eqref{E:uTtoK} and converting to the Landau frame, we find that the stress tensor takes the form
\begin{equation}
    T_s^{\mu\nu}
      =P_0 \left(4 u^{\mu}u^{\nu} + g^{\mu\nu}\right)
       + \lambda_3 
        \Omega^{\alpha\langle\mu}\Omega_\alpha{}^{\nu\rangle}
	+ \kappa_g \left(R^{\langle \mu\nu\rangle } - 2  u_{\alpha}R^{\alpha\langle \mu\nu \rangle\beta}u_{\beta} \right)       
       +\left(\substack{\hbox{parity odd} \\ \hbox{terms} }\right) + \mathcal{O}(\nabla^3)\,.
\end{equation}
Here the parity odd terms come from shifting the first order shear tensor from the hydrostatic frame to the Landau frame \cite{BhattacharyyaDavidThakur:2013}, and we have defined\footnote{We have used that for a conformal theory in 3+1 dimensions $T\frac{\partial \kappa_g}{\partial T}+\mu \frac{\partial \kappa_g}{\partial \mu}=2 \kappa_g$. This follows from noting that in this case $\kappa_g=T^2 F\left(\frac{\mu}{T} \right)$ with $ F\left(\frac{\mu}{T} \right)$ some function.}
\begin{equation}
\label{E:lambda3conventional}
	\lambda_3 = -4 \lambda_3^h - 2 \kappa_g + \frac{\left(4 c_A \mu^3 + 32 \pi^2 c_m T^2 \mu \right)^2}{4 P_0} \,.
\end{equation}

Let us now extend this result to hydrostatics with spin currents. To describe spin hydrodynamics in hydrostatic equilibrium, we temporarily turn on a background contorsion field $K^{\mu}{}_{\nu\rho}$. In this case, we add the following term to \eqref{E:timeindependence}:
\begin{equation}
	\pounds_V\omega_\mu{}^{ab}+D_\mu\theta_V{}^{ab}=0\,,
\end{equation}
and the following term to \eqref{E:uTtoK}:
\begin{equation}
	\frac{\mu^{ab}}{T}= V^\mu\omega_\mu{}^{ab}+\theta_V{}^{ab}
\end{equation}
Analogously to \eqref{E:hconstraints}, we have the equilibrium relations
\begin{equation}
\label{E:hydrostatic2}
	M^{\mu\nu} = -\Omega^{\mu\nu} +\mathcal K_A^{\mu\nu} +k_A^{\mu\nu}\,,
	\qquad
	m^\mu = a^\mu+\frac{1}{3}k_V^\mu-\mathcal K_V^\mu\,,
\end{equation}
where $\mathcal{K}_A$, $k_A$ and $\mathcal{K}_V$ are components of the contorsion tensor described in appendix \ref{A:contorsion}.
See \cite{Gallegos:2022jow}.\footnote{Note that using \eqref{E:zeroEOM} one can recast \eqref{E:hydrostatic2} in the absence of torsion in the form
$
	\frac{\mu^{\alpha\beta}}{T} = -\nabla^{[\alpha}\frac{u^{\beta]}}{T}\,.
$
The term on the right hand side of the last equation is referred to as thermal vorticity \cite{Becattini:2013fla}.}

The hydrostatic partition function $W$ in the presence of contorsion extends the partition function in \eqref{E:WnoK}. Naively, we write 
\begin{equation}
	W=W_0 - \int \rho_M M_{\alpha\beta}M^{\alpha\beta}  \sqrt{-g}  d^4x + W_K  \,.
\end{equation}
The second term generates the spin density, while the last denotes a contribution that vanishes when the contorsion is set to zero. Only terms linear in the contorsion can contribute to the constitutive relations after $K^{\mu}{}_{\nu\rho}$ is set to zero, and these generate improvement pairs $(J^{\alpha\mu\nu},B^{\mu\nu}[J])$ whose contribution to the manifestly symmetric stress tensor vanishes \cite{Gallegos:2022jow}.

Now, using \eqref{E:hydrostatic2}, we find
\begin{equation}
	M^{\mu\nu}M_{\mu\nu} = \Omega^{\mu\nu}\Omega_{\mu\nu} + 2 \mathcal K_A^{\mu\nu} M_{\mu\nu} + 2k_A^{\mu\nu}M_{\mu\nu} + \mathcal{O}(K^2)\,.
\end{equation}
Thus,
\begin{multline}
	W = \int P(T,\mu) \sqrt{-g} d^4x + W_A 
	\\
	+ \int \left( \left(\lambda_3^h-\rho_M\right)  \,\Omega^{\alpha\beta}\Omega_{\alpha\beta} -\frac{\kappa_g}{2}  (R+6 a_{\alpha}a^{\alpha} ) +3 \frac{\partial \kappa_g}{\partial \mu} E_{\alpha}a^{\alpha} + \ldots \right) \sqrt{-g} d^4x + W_K\,.
\end{multline}
The torsionless constitutive relations depend only on the combination $X = \lambda_3^h-\rho_M$. In the spin-hydrodynamic description of the fluid, one relabels $(\lambda_3^h,\rho_M)\to(0,-X)$, whereas in the conventional description using a manifestly symmetric stress tensor, one takes $(\lambda_3^h,\rho_M)\to(X,0)$. Since the conventional hydrodynamic description leads to \eqref{E:lambda3conventional}, we obtain \eqref{E:rhoMlambda3}.

\section{Discussion}
\label{S:discussion}

In this work, we focused on the spin density, $\rho^{\mu\nu}$, which controls the ``ideal'' constitutive relations for the spin current (in the hydrostatic frame)
\begin{equation}
	S^{\mu\nu\rho} = u^{\mu} \rho^{\nu\rho}\,.
\end{equation}
We have shown that in a conformal field theory and to first order in derivatives the spin density is controlled by a single transport coefficient $\rho_M(T,\mu)$ via
\begin{equation}
	\rho^{\mu\nu} = -4 \rho_M M^{\mu\nu}\,.
\end{equation}
The main result of this work, captured by equation \eqref{E:rhoMlambda3}, is the evaluation of $\rho_M$ using the anomaly coefficients $c_A$ and $c_m$ as well as second order  hydrostatic data $\lambda_3$ and $\kappa$ of the standard, manifestly symmetric fluid stress tensor, as described in, e.g., \cite{BhattacharyyaHubenyMinwallaRangamani2008,BaierRomatschkeSonStarinetsStephanov2008}. 

The second order transport coefficients $\lambda_3$ and $\kappa$ have been computed holographically in \cite{BhattacharyyaHubenyMinwallaRangamani2008,BaierRomatschkeSonStarinetsStephanov2008,Erdmenger:2008rm,Banerjee:2008th,ArnoldVamanWuXiao:2011,SaremiSohrabi:2011,Grozdanov:2014kva,GrieningerShukla:2021}, at weak coupling or in free quantum field theories in \cite{MooreSohrabi:2010,MooreSohrabi:2012,KovtunShukla:2018,Shukla:2019}, and, for $\kappa$, using lattice gauge theory in \cite{PhilipsenSchaefer:2014}.
Table \ref{T:various} collects the resulting expressions for the spin density in the relevant theories. Details on each entry can be found in Appendix \ref{A:rhoMtable}.
\begin{table}[h]
\centering
\scriptsize
\begingroup
\renewcommand{\arraystretch}{2}
\resizebox{\textwidth}{!}{%
\begin{tabular}{|l|c|c|c|c|c|c|}
\hline
Theory & $\kappa_g$ & $\lambda_3$ & $c_A$ & $c_m$ & $\rho_M$ & Reference \\
\hline
Real conformal scalar
 & $0$ & $\frac{T^2}{18}$ & $0$ & $0$ & $\frac{T^2}{72}$ & \cite{KovtunShukla:2018} \\
\hline
Massless Dirac fermion, $\mu=0$
 & $\frac{T^2}{72}$ & $0$ & $0$ & $0$ & $\frac{T^2}{144}$ & \cite{KovtunShukla:2018} \\
\hline
Massless Dirac fermion, $T=0$
 & $\frac{\mu_V^2}{24\pi^2}$ & $0$ & $0$ & $0$ & $\frac{\mu_V^2}{48\pi^2}$ & \cite{Shukla:2019} \\
\hline
Maxwell field
 & $\frac{T^2}{18}$ & $-\frac{T^2}{3}$ & $0$ & $0$ & $-\frac{T^2}{18}$ & \cite{KovtunShukla:2018} \\
\hline
$\mathcal N=4$ $SU(N_c)$ SYM, $\mu=0$
 & $\frac{N_c^2T^2}{8}$ & $0$ & 0 & 0 & $\frac{N_c^2T^2}{16}$ & \cite{BaierRomatschkeSonStarinetsStephanov2008,BhattacharyyaHubenyMinwallaRangamani2008} \\
\hline
$\mathcal N=4$ $SU(N_c)$ SYM, $\mu\neq0$
 & $\frac{N_c^2r_+^2}{8\pi^2}$
 & $\frac{N_c^2\ r_+^2 \mu^2}{\pi^2(6 r_+^2+2\mu^2)}$
 & $\left|c_A\right|=\frac{N_c^2}{24\sqrt{3}\pi^2}$
 & $0$ & $\rho_M^{\rm EM}$
 & \cite{Erdmenger:2008rm,Banerjee:2008th,GrieningerShukla:2021} \\
\hline
$\mathcal N=2$ $Sp(N_c)$ theory, $\mu=0$
 & $T^2\left(\frac{N_c^2}{4}+\frac{N_c}{4}+\mathcal O(1)\right)$
 & $T^2\left(-\frac{7N_c}{8}+\mathcal O(1)\right)$
 & 0 & 0
 & $T^2\left(\frac{N_c^2}{8}-\frac{3N_c}{32}+\mathcal O(1)\right)$
 & \cite{Blau:1999vz,Kats:2007mq,Myers:2010tj,GrozdanovStarinets:2017} \\
\hline
\end{tabular}%
}
\endgroup
\caption{Values for the spin density, $\rho_M$, obtained from \eqref{E:rhoMlambda3} for a variety of conformal field theories. In the holographic  Einstein--Maxwell model we used 
$
	r_+ = \frac{\pi T}{2}\left(1+q \right)
$
with
$
	q= \sqrt{1+\frac{2\mu^2}{3\pi^2T^2}}
$
and
$
 \rho_M^{\rm EM}  =\frac{N_c^2T^2}{64}\frac{11-51q+57q^2-9q^3}{3q-1}
$.
The $Sp(N_c)$ results are given through order $N_c$ in the large-$N_c$ expansion. See Appendix \ref{A:rhoMtable} for details.
 }
\label{T:various}
\end{table}

Given that holographic results have successfully captured properties of the quark gluon plasma \cite{PolicastroSonStarinets2001,CasalderreySolana:2011us}, it may not be unreasonable to use the results for $\mathcal{N}=4$ SYM with $N_c=3$, which gives 
\begin{equation}
	\rho_M^{\mathcal{N}=4,\,\infty} = \frac{9}{16} T^2
\end{equation}
as a preliminary estimate for spin density in heavy ion collisions \cite{Gallegos:2021bzp,Becattini:2020ngo}. 

We note that at zero coupling, an $\mathcal N=4$ SYM vector multiplet contains six real conformal scalars, four Weyl fermions (equivalently two Dirac fermions), and one gauge field, all in the adjoint representation. Hence, for $\mathcal{N}=4$ super Yang-Mills at zero coupling, we find
\begin{equation}
\label{E:N4zero}
	\rho_M^{\mathcal N=4,\,0}
	=\frac{N_c^2-1}{24}T^2\,.
\end{equation}
We can compare \eqref{E:N4zero} to massless QCD at zero coupling. To this end, consider an $SU(N_c)$ gauge theory with $N_f$ massless Dirac fermions in the fundamental representation of $SU(N_c)$. There are $N_cN_f$ Dirac fermions and $N_c^2-1$ gauge fields. Adding their free-field contributions from Table \ref{T:various} gives
\begin{equation}
	\rho_M^{\mathrm{QCD},\,0}
	=\left(\frac{N_cN_f}{144}-\frac{N_c^2-1}{18}\right)T^2\,.
\end{equation}
For a benchmark relevant to the quark--gluon plasma, we take $N_c=3$ and $N_f=3$, corresponding to the usual $2+1$ flavor matter content, with the strange quark treated as massless \cite{Bazavov:2014pvz}. This gives
\begin{equation}
	\rho_M^{\mathrm{QCD},\,0}\big|_{N_c=N_f=3}
	=-\frac{55}{144}T^2\,.
\end{equation}

Extensions of the holographic results in Table \ref{T:various} to theories with a mixed gauge-gravitational anomaly or other four derivative perturbations of Einstein theory (such as the one described in \cite{Cremonini:2009sy}) are demanding  but computable. 

With relatively little additional work, these results could also be extended to non conformal  theories where the charge density (in $3+1$ dimensions) is controlled by additional $\rho_m$ and $\rho_{\tilde{M}}$ terms (see Appendix \ref{A:eqConst}). These terms are associated with $m^{\mu}m_{\mu}$ and $m^{\mu}\tilde{M}_{\mu}$ contributions to the partition function which can be shifted to more standard $a_{\mu}a^{\mu}$ and $a_{\mu}\tilde{\Omega}^{\mu}$ contributions. It would be interesting to work out a general expression for the spin density of nonconformal field theories using this approach.

Determining non-hydrostatic spin related transport seems more challenging but, as in the hydrostatic case, it is possible that most spin related transport is computable from already known second order hydrodynamic transport coefficients. We hope to explore this possibility in the future.

\begin{appendix}

\section{Deriving the constitutive relations for hydrodynamics with spin}
\label{A:lightningreview}
In this section we sketch out the derivation of hydrodynamics with a spin current. We first compute the hydrostatic contributions to the constitutive relations, then add non hydrostatic terms, and end with an entropy analysis.

\subsection{Hydrostatics}
\label{A:eqConst}
As discussed in \cite{Gallegos:2022jow}, the most general parity violating hydrostatic pressure term in $d=3+1$ dimensions can be described by a function $P=P\left(T,\mu,m_{\left(0\right)},M_{\left(1\right)},\tilde{M}\right)$
 , where $m_{\left(0\right)}=m_{a}m^{a}$, $M_{\left(1\right)}=M_{ab}M^{ba}$ and $\tilde{M}=\varepsilon^{abcd}u_{a}m_{b}M_{cd}=\frac{1}{4}\eps^{abcd}\mu_{ab}\mu_{cd}$.  Expanding $P$ to second order in derivatives we find
 \begin{equation}
 	    P=P_{0}\left(T,\mu\right)+\rho_{M}M_{\left(1\right)}+\rho_{m}m_{\left(0\right)}+
    \rho_{\tilde{M}}\tilde{M}+O\left(\nabla^{4}\right)
 \end{equation}
 Varying $W_{id} = \int P \sqrt{-g} d^4x$ with respect to the metric, gauge field and contorsion one obtains the constitutive relations in \eqref{E:idstresscurrent} with

\begin{align}
    \rho_{ab}&=2\frac{\partial P}{\partial\mu^{ab}}=-4\rho_{M}M_{ab}-4\rho_{m}u_{[a}m_{b]}+2\rho_{\tilde{M}}\left(u_{[a}\tilde{M}_{b]}+\tilde{m}_{ab}\right)+O\left(\nabla^{3}\right)\,, \label{termRel-2}\\
    P_{a}&=P_a{}^b\frac{\partial P}{\partial u^{b}}=-\left(4\rho_{M}-2\rho_{m}\right)m_{b}\tud Mba+O\left(\nabla^{4}\right)\,, \label{termRel-3}
    \\
    \begin{split}
    \eps&=TS+\mu\rho+\frac{1}{2}\mu^{ab}\rho_{ab}-P
    \\&=\eps\big|_{\mu^{ab}=0}+\left(\rho_{M}+T\frac{\partial\rho_{M}}{\partial T}+\mu\frac{\partial\rho_{M}}{\partial\mu}\right)M_{\left(1\right)}+\left(\rho_{m}+T\frac{\partial\rho_{m}}{\partial T}+\mu\frac{\partial\rho_{m}}{\partial\mu}\right)m_{\left(0\right)}\\
        &+\left(\rho_{\tilde{M}}+T\frac{\partial\rho_{\tilde{M}}}{\partial T}+\mu\frac{\partial\rho_{\tilde{M}}}{\partial\mu}\right)\tilde{M}+O\left(\nabla^4\right)\,. \label{termRel-4}
    \end{split}
\end{align}
In a conformal theory $\rho_m=0$ and $\rho_{\tilde{M}}=0$ leading to \eqref{E:idealexpansion}.

To obtain the most general hydrostatic constitutive relations we expand the generating function $W$ in a derivative expansion. To this end, we must classify all possible scalars  (and pseudoscalars) which may contribute to the constitutive relations at the appropriate order in derivatives. For the case at hand we would like to expand the spin current, charge current and symmetric component of the stress tensor to first order in derivatives and the antisymmetric component of the stress tensor to second order in derivatives. Thus, we must classify all possible first order scalar (and pseudoscalar) expressions and all possible second scalars which are linear in contorsion (see \cite{Gallegos:2021bzp,Gallegos:2022jow}). We have tabulated all such terms in tables \ref{T:scalars} and \ref{T:Pscalars}.
\begin{table}[hbt]
\begin{center}
\begin{tabular}{|c|c|c|ccc|}
\hline 
0th order & \multicolumn{5}{c|}{None}\tabularnewline
\hline 
1st order & \multicolumn{2}{c|}{Conformal contorsion linear} & \multicolumn{3}{c|}{$\kappa$}\tabularnewline
\hline 
\multirow{5}{*}{2nd order} & \multirow{2}{*}{Contorsion independent} & Conformal & \multicolumn{3}{c|}{
$s_{\left(1\right)} =B_{ab}M^{ab}$
}\tabularnewline
\cline{3-6} 
 &  & Non-conformal & 
 \multicolumn{3}{c|}{$s_{\left(2\right)}=E_{a}m^{a}$}\tabularnewline
\cline{2-6} 
 & 
 \multirow{3}{*}{Contorsion linear} & 
 \multirow{2}{*}{Conformal} & $t_{\left(1\right)}=M_{ab}k_{A}^{ab}$ & $t_{\left(3\right)}=E_{a}k_{V}^{a}$ & $t_{\left(5\right)}=B_{ab}k_{A}^{ab}$\tabularnewline
 &  &  & $t_{\left(2\right)}=M_{ab}\mathcal{K}_{A}^{ab}$ & $t_{\left(4\right)}=E_{a}\mathcal{K}_{V}^{a}$ & $t_{\left(6\right)}=B_{ab}\mathcal{K}_{A}^{ab}$\tabularnewline
\cline{3-6} 
 &  & Non-conformal & 
 \multicolumn{3}{c|}{$t_{\left(7\right)}=m_{a}k_{V}^{a}$ $\qquad$$t_{\left(8\right)}=m_{a}\mathcal{K}_{V}^{a}$}\tabularnewline
\hline 
\end{tabular}
\caption{All scalar expressions that contribute to the non ideal constitutive relations in the presence of torsion and U(1) fields. The electric and magnetic fields are given by $E_\mu=F_{\mu\nu}u^\nu$ and $B_{\mu\nu}=\frac{1}{2}\tdu{P}{\mu}{\rho}\tdu{P}{\nu}{\sigma}F_{\rho\sigma}$.
\label{T:scalars}}
\end{center}
\end{table}
\begin{table}[hbt]
\begin{center}
\begin{tabular}{|c|c|c|ccc|}
\hline 
0th order & \multicolumn{5}{c|}{None}\tabularnewline
\hline 
1st order & \multicolumn{2}{c|}{Conformal contorsion linear} & \multicolumn{3}{c|}{$\tilde{\kappa}$}\tabularnewline
\hline 
\multirow{5}{*}{2nd order} & \multirow{2}{*}{Contorsion independent} & Conformal & \multicolumn{3}{c|}{
$\tilde{s}_{\left(1\right)} =E_{a}\tilde{M}^{a}$
}\tabularnewline
\cline{3-6} 
 &  & Non-conformal & 
 \multicolumn{3}{c|}{$\tilde{s}_{\left(2\right)}=\tilde{B}_{a}m^{a}$}\tabularnewline
\cline{2-6} 
 & 
 \multirow{3}{*}{Contorsion linear} & 
 \multirow{2}{*}{Conformal} & $\tilde{t}_{\left(1\right)}=\tilde{M}_{a}k_{V}^{a}$ & $\tilde{t}_{\left(3\right)}=\tilde{B}_{a}k_{V}^{a}$ & $\tilde{t}_{\left(5\right)}=\tilde{E}_{ab}k_{A}^{ab}$\tabularnewline
 &  &  & $\tilde{t}_{\left(2\right)}=\tilde{M}_{a}\mathcal{K}_{V}^{a}$ & $\tilde{t}_{\left(4\right)}=\tilde{B}_{a}\mathcal{K}_V^{a}$ & $\tilde{t}_{\left(6\right)}=\tilde{E}_{ab}\mathcal{K}_{A}^{ab}$\tabularnewline
\cline{3-6} 
 &  & Non-conformal & 
 \multicolumn{3}{c|}{$\tilde{t}_{\left(7\right)}=\tilde{m}_{ab}k_{A}^{ab}$ $\qquad$$\tilde{t}_{\left(8\right)}=\tilde{m}_{ab}\mathcal{K}_{A}^{ab}$}\tabularnewline
\hline 
\end{tabular}
\caption{All pseudoscalar expressions that contribute to the non ideal constitutive relations in the presence of torsion and U(1) fields. The electric and magnetic fields are given by $E_\mu=F_{\mu\nu}u^\nu$ and $B_{\mu\nu}=\frac{1}{2}\tdu{P}{\mu}{\rho}\tdu{P}{\nu}{\sigma}F_{\rho\sigma}$.
\label{T:Pscalars}}
\end{center}
\end{table}

With tables \ref{T:scalars} and \ref{T:Pscalars} at hand, the partition function takes the form
\begin{equation}\label{E:PartitionFunction}
W = \intop d^4x \left|e\right|\left(P+\mathcal{W}_A+\mathcal{W}_{(1)} +\mathcal{W}_{(2)}+ O\left(\nabla^3\right)\right)\,,
\end{equation}
where
\begin{align}
\begin{split}
	\mathcal{W}_{\left(1\right)}&=\zeta^{\left(1\right)}\kappa+\tilde{\zeta}^{\left(1\right)}\tilde{\kappa}\,,
	\\
	\mathcal{W}_{\left(2\right)}&=\xi_{i}^{\left(2\right)}s_{\left(i\right)}+\tilde{\xi}_{i}^{\left(2\right)}\tilde{s}_{\left(i\right)}+\zeta_{i}^{\left(2\right)}t_{\left(i\right)}+\tilde{\zeta}_{i}^{\left(2\right)}\tilde{t}_{\left(i\right)}\,,
\end{split}
\end{align}
and $W_A$ is such that it encodes the appropriate anomalous constitutive relations. See \cite{Jensen:2012kj,Jensen:2013kka,Jensen:2013rga}.

Varying $W$ with respect to the metric, gauge field and contorsion, and setting the contorsion to zero, one finds
\begin{align}
	T^{\mu\nu}_h& = e^{\nu a} \frac{1}{|e|}  \frac{\delta W}{\delta e_\mu{}^a} +P_{BZ}^{\mu\nu} 
		= T_{id}^{\mu\nu}+T_{A}^{\mu\nu}+T_{hnBR}^{\mu\nu}+\frac{1}{2}\nabla_{\lambda}\left(S_{hBR}^{\lambda\mu\nu}-S_{hBR}^{\mu\lambda\nu}-S_{hBR}^{\nu\lambda\mu}\right)\,,\\
	J^{\mu}_h&=\frac{1}{\sqrt{-g}} \frac{\delta W}{\delta A_{\mu}} + P^{\mu}_{BZ} 
		= J_{id}^{\mu}+J_{A}^{\mu}+O(\nabla^2)\,,\\
	S^{\lambda\mu\nu}_h&= e^\mu{}_a e^\nu{}_b \frac{2}{|e|} \frac{\delta W}{\delta K_{\lambda ab}} 
		=S_{id}^{\lambda\mu\nu}+S_{hnBR}^{\lambda\mu\nu}+S_{hBR}^{\lambda\mu\nu}\,,
\end{align}
where $T_{id}^{\mu \nu}$, $J_{id}^{\mu}$ and $S^{\lambda \mu \nu}_{id}$ are given by \eqref{E:idstresscurrent} with  \eqref{termRel-2}-\eqref{termRel-4} and $P_{BZ}^{\mu}$ and $P_{BZ}^{\mu\nu}$ are the Bardeen--Zumino improvement terms \cite{BardeenZumino:1984,Jensen:2013kka}.
The (covariant) anomalous terms $T^{\mu \nu}_A$ and $J_A^{\mu}$ are  given by
\begin{align}
\begin{split}
	T_A^{\mu\nu}=&-2\left(3c_A\mu^2+8\pi^2c_mT^2\right) u^{(\mu}\tilde B^{\nu)}
	-2\left(2c_A\mu^3+16\pi^2c_m\mu T^2\right) u^{(\mu}\tilde\Omega^{\nu)}\,, \\
         J_A^\mu=&-6c_A\mu\,\tilde B^\mu -\left(3c_A\mu^2+8\pi^2c_mT^2\right) \tilde\Omega^\mu\,.
\end{split}
\end{align}
	
The term $S^{\mu \nu}_{hBR}$ is a pure improvement term given by 
\begin{align}
    \begin{split}
    S_{hBR}^{\lambda\mu\nu}&=2\zeta^{\left(1\right)}g^{\lambda[\mu}u^{\nu]}+\frac{1}{3}\tilde{\zeta}^{\left(1\right)}\eps^{\lambda\mu\nu\rho}u_{\rho}
    \\
    &+2\zeta_{1}^{\left(2\right)}M^{[\mu\nu}u^{\lambda]}+2\zeta_{2}^{\left(2\right)}\left(M^{\mu\nu}u^{\lambda}-M^{[\mu\nu}u^{\lambda]}\right)
    \\&+2\tilde{\zeta}_{1}^{\left(2\right)}g^{\lambda[\mu}\tilde{M}^{\nu]}+2\tilde{\zeta}_{2}^{\left(2\right)}\left(u^{\lambda}u^{[\mu}\tilde{M}^{\nu]}+\frac{1}{3}g^{\lambda[\mu}\tilde{M}^{\nu]}\right)
    \\&+2\zeta_{3}^{\left(2\right)}g^{\lambda[\mu}E^{\nu]}+2\zeta_{4}^{\left(2\right)}\left(u^{\lambda}u^{[\mu}E^{\nu]}+\frac{1}{3}g^{\lambda[\mu}E^{\nu]}\right)
    \\&+2\zeta_{5}^{\left(2\right)}B^{[\mu\nu}u^{\lambda]}+2\zeta_{6}^{\left(2\right)}\left(B^{\mu\nu}u^{\lambda}-B^{[\mu\nu}u^{\lambda]}\right)
    \\&+2\tilde{\zeta}_{3}^{\left(2\right)}g^{\lambda[\mu}\tilde{B}^{\nu]}+2\tilde{\zeta}_{4}^{\left(2\right)}\left(u^{\lambda}u^{[\mu}\tilde{B}^{\nu]}+\frac{1}{3}g^{\lambda[\mu}\tilde{B}^{\nu]}\right)
    \\&+2\tilde{\zeta}_{5}^{\left(2\right)}\tilde{E}^{[\mu\nu}u^{\lambda]}+2\tilde{\zeta}_{6}^{\left(2\right)}\left(\tilde{E}^{\mu\nu}u^{\lambda}-\tilde{E}^{[\mu\nu}u^{\lambda]}\right)\\&+2\zeta_{7}^{\left(2\right)}g^{\lambda[\mu}m^{\nu]}+2\zeta_{8}^{\left(2\right)}\left(u^{\lambda}u^{[\mu}m^{\nu]}+\frac{1}{3}g^{\lambda[\mu}m^{\nu]}\right)
    \\&+2\tilde{\zeta}_{7}^{\left(2\right)}\tilde{m}^{[\mu\nu}u^{\lambda]}+2\tilde{\zeta}_{8}^{\left(2\right)}\left(\tilde{m}^{\mu\nu}u^{\lambda}-\tilde{m}^{[\mu\nu}u^{\lambda]}\right)+O\left(\nabla^2\right)\,.
    \end{split}
\end{align}
The remaining terms $T^{\mu \nu}_{hnBR}$ and $S^{\lambda \mu \nu}_{hnBR}$ are hydrostatic non-ideal and non-improvement terms given by 
\begin{align}
    \begin{split}
    S_{hnBR}^{\lambda\mu\nu}&=2\xi_{1}^{\left(2\right)}u^{\lambda}B^{\mu\nu}-2\tilde{\xi}_{1}^{\left(2\right)}u^{\lambda}\tilde{E}^{\mu\nu}\\&-2\xi_{2}^{\left(2\right)}u^{\lambda}u^{[\mu}E^{\nu]}-2\tilde{\xi}_{2}^{\left(2\right)}u^{\lambda}u^{[\mu}\tilde{B}^{\nu]}+O\left(\nabla^2\right)\,,
    \end{split}\\
    \begin{split}
    T_{hnBR}^{[\mu\nu]}&=\xi_{1}^{\left(2\right)}\left(2m_{\rho}B^{\rho[\nu}u^{\mu]}+2M^{\rho[\nu}\tdu B{\rho}{\mu]}\right)
    \\&+\tilde{\xi}_{1}^{\left(2\right)}\left(2u^{[\mu}\left(2\tud M{\nu]}{\rho}\tilde{B}^{\rho}-\tilde{m}^{\nu]\rho}E_{\rho}\right)+\tilde{M}^{[\mu}E^{\nu]}\right)
    \\&+\xi_{2}^{\left(2\right)}\left(\left(4m_{\rho}B^{\rho[\nu}+E_{\rho}M^{\rho[\nu}\right)u^{\mu]}-m^{[\nu}E^{\mu]}\right)
    \\&-\tilde{\xi}_{2}^{\left(2\right)}\left(m^{[\mu}\tilde{B}^{\nu]}+u^{[\mu}\tud M{\nu]}{\rho}\tilde{B}^{\rho}\right)+O\left(\nabla^3\right)\,,
    \end{split}
    \\T_{hnBR}^{(\mu\nu)}&=O\left(\nabla^2\right)\,.
\end{align}
In the conformal case and in the absence of external electromagnetic fields, the hydrostatic constitutive relations reduce to the hydrostatic part of the full constitutive relations shown in \eqref{E:fullconstitutive}.

\subsection{Non-equilibrium constitutive relations}
\label{A:nonhydrostatic}
The full constitutive relations for the currents $T^{\mu \nu}$, $S^{\lambda \mu \nu}$ and $J^\mu$ are
\begin{align}
    T^{\mu \nu} &= T^{\mu \nu}_h + T^{\mu \nu}_{nh} +B^{\mu\nu}[\Phi_{nh}] \, , \\
    S^{\lambda \mu \nu} &= S^{\lambda \mu \nu}_h + S^{\lambda \mu \nu}_{nh} + \Phi_{nh}^{\lambda\mu\nu}\, , \\
    J^\mu &= J_h^\mu + J^\mu_{nh} \, ,
\end{align} 
Where $T^{\mu \nu}_h$, $S^{\lambda \mu \nu}_h$ and $J^\mu_h$ are the hydrostatic currents obtained in appendix \ref{A:eqConst} and $T^{\mu \nu}_{nh}$, $S^{\lambda \mu \nu}_{nh}$ and  $J^\mu_{nh}$ are on-shell inequivalent tensors which vanish in hydrostatic equilibrium.\footnote{Note that we are taking $T^{\mu \nu}_h$ such that it contains the necessary non-hydrostatic contributions so that the corresponding contributions from $S^{\lambda \mu \nu}_{hBR}$ are kept as a improvement terms. Namely, any non-hydrostatic contribution that follows from $\frac{1}{2}\nabla_\lambda \left(S^{\lambda \mu \nu}_{h BR} - S^{\mu \lambda \nu}_{hBR} - S^{\nu \lambda \mu}_{h BR} \right)$ is kept in the constitutive relations for $T^{\mu \nu}_h$.}
To write the constitutive relations for the non hydrostatic components explicitly, we need to list all possible on-shell inequivalent non equilibrium tensor structures up to first in derivatives. (In principle, we need to include second order in derivative antisymmetric tensors but, as we will argue (and as discussed in section \ref{SS:frames}), we choose a fluid frame where all such tensor structures do not contribute to the constitutive relations except for pure improvement terms.) We have listed all such tensor structures in table \ref{T:nhTensors}, where $A^a$ and $\mathcal{E}^a$ are shorthand for
\begin{align}
\begin{split}
    A^a&= P^{ab}\nabla_b T + T a^a\,,
    \\
    \mathcal{E}^{a}&=E^{a}-TP^{ab}\nabla_{b}\frac{\mu}{T}\,.
\end{split}
\end{align}
\begin{table}[h!]
    \centering
    \begin{tabular}{|c|c|c|c|}
        \hline
        \textbf{Tensor type} & \textbf{Non-hydrostatic data} & \textbf{EOM} & \textbf{Independent data} \\ \hline
        Spin 0 & $\dot{T},\dot{\mu},\nabla_\alpha u^\alpha$ & \begin{tabular}{c}
            $u_{\nu}\left(\nabla_{\mu}T_{C}^{\mu\nu}+F^{\mu\nu}J_{\mu}\right)=0$ \\[5pt]
            $\nabla_{\mu}J^{\mu}-\frac{3}{4}c_{A}\eps^{\mu\nu\rho\sigma}F_{\mu\nu}F_{\rho\sigma}=0$
        \end{tabular} & $\nabla_\alpha u^\alpha$ \\ \hline
        Spin 1 & $A^{a},\mathcal{E}^{a},\hat{m}^{a},\hat{M}^{ab}$ & \begin{tabular}{c}
            $\tud P{\nu}{\rho}\left(\nabla_{\mu}T_{C}^{\left(\mu\rho\right)}+F^{\mu\rho}J_{\mu}\right)=0$ \\[5pt]
            $u_{\mu}T_{C}^{\left[\mu\nu\right]}=0$ \\[5pt]
            $\tud P{\mu}{\rho}\tud P{\nu}{\rho}T_{C}^{\left[\rho\sigma\right]}=0$
        \end{tabular} & $\mathcal{E}^a$ \\ \hline
        Spin 2 & $\sigma^{ab}$ & None & $\sigma^{ab}$ \\ \hline
    \end{tabular}
    \caption{A tabulation of all non-hydrostatic first order tensors of the residual $SO(3)$ symmetry associated with $u^{\mu}$. Pseudo tensors can be obtained by acting with the epsilon tensor on these. See \eqref{E:hodge}.}
    \label{T:nhTensors}
\end{table}

To construct $T^{(\mu \nu)}_{nh}$ and $J^\mu_{nh}$ we use fix the fluid frame for the non hydrostatic components to the Landau frame where
\begin{align}
    u_\mu T^{(\mu \nu)}_{nh} &=0 \, , \\ 
    u_\mu J^\mu_{nh} &= 0 \, .
\end{align}
Thus, 
\begin{align}
    T_{nh}^{(\mu \nu)}&= - \eta \sigma^{\mu \nu} - \zeta P^{\mu\nu} \nabla_\alpha u^\alpha \, , \\
    J^\mu_{nh}&= \sigma_E \mathcal{E}^\mu\,.
\end{align}
\begin{align}\label{const:Anti:EnergyMom}
T_{nh}^{\left[\mu\nu\right]}&=-\sigma_{\mathcal{E}}\mathcal{E}^{[\mu}u^{\nu]}-\sigma_{\tilde{\mathcal{E}}}\tilde{\mathcal{E}}^{\mu\nu}+\left(\sigma_{m}\hat{m}^{[\mu}-\frac{1}{2}\sigma_{\tilde{M}}\tilde{\hat{M}}^{[\mu}\right)u^{\nu]}+\sigma_{M}\hat{M}^{\mu\nu}+\sigma_{\tilde{m}}\tilde{\hat{m}}^{\mu\nu}\,.    
\end{align}
Note that a frame choice, previously introduced in \cite{Gallegos:2022jow}, has been  used in \eqref{const:Anti:EnergyMom} so that only $\mathcal{O}\left(\nabla\right)$ terms are contained in $T_{nh}^{[\mu \nu]}$, this corresponds to a fixing a frame such that all higher order contributions are absorbed into the definition of the non-equilibrium spin potentials $m^\mu$ and $ M^{\mu \nu}$. Therefore, the antisymmetric stress tensor is characterized by six non-equilibrium transport coefficients, $\sigma_m$, $\sigma_M$, $\sigma_{\tilde M}$, $\sigma_{\tilde m}$, $\sigma_{\mathcal{E}}$ and $\sigma_{\tilde{ \mathcal{E}}}$ appearing in \eqref{const:Anti:EnergyMom} and pure improvement terms which we will discuss shortly.

For the spin current we find that there are eight possible tensor structures which we write in the form
\begin{align}
    \begin{split}
    S_{nh}^{\lambda\mu\nu}&=2\sigma_{1}\sigma^{\lambda[\mu}u^{\nu]}+2\tilde{\sigma}_{1}\tud{\eps}{\mu\nu}{\rho\sigma}\sigma^{\lambda\rho}u^{\sigma}
    \\&-2\sigma_{2}g^{\lambda[\mu}\mathcal{E}^{\nu]}-2\sigma_{3}\left(u^{\lambda}u^{[\mu}\mathcal{E}^{\nu]}+\frac{1}{3}g^{\lambda[\mu}\mathcal{E}^{\nu]}\right)
    \\&-2\tilde{\sigma}_{2}u^{[\lambda}\tilde{\mathcal{E}}^{\mu\nu]}-2\tilde{\sigma}_{3}\left(u^{\lambda}\tilde{\mathcal{E}}^{\mu\nu}-u^{[\lambda}\tilde{\mathcal{E}}^{\mu\nu]}\right)
    \\&+2\sigma_{4}\left(\nabla_\alpha u^\alpha\right) g^{\lambda[\mu}u^{\nu]}+\frac{1}{3}\tilde{\sigma}_{4}\left(\nabla_\alpha u^\alpha\right)\eps^{\lambda\mu\nu\rho}u_{\rho}
    +O\left(\nabla^2\right)\,.
\end{split}
\end{align}
In addition, there are eight parallel tensor structures that we contribute to $\Phi^{\lambda\mu\nu}_{nh}$ which we will not write explicitly. 
In the conformal case and in the absence of external electromagnetic fields, the non-hydrostatic constitutive relations reduce to the non-hydrostatic parts of the full constitutive relations shown in \eqref{E:fullconstitutive}.

\subsection{Entropy current}
\label{A:entropyII}

In sections \ref{A:eqConst} and \ref{A:nonhydrostatic} we classified all possible allowed constitutive relations compatible with thermodynamic equilibrium. An additional constraint that we have not imposed is that a local version of the second law of thermodynamics holds \cite{LandauLifshitz:1987,GloriosoLiu:2016,JensenMarjiehPinzaniFokeevaYarom:2018}. More formally, one posits that there exists an entropy current $J_s^{\mu}$ satisfying 
\begin{equation}
\label{E:local2nd}
	\nabla_{\mu}J_s^{\mu} \geq 0
\end{equation}
under the equations of motion,
 and that its constitutive relations are 
 \begin{equation}
\label{E:constitutive2nd}
	J_s^{\mu} = s u^{\mu} + \mathcal{O}(\nabla)
\end{equation}
where $s=dP/dT$ as in \eqref{E:firstlaw}. 
The relations \eqref{E:local2nd} and \eqref{E:constitutive2nd} turn out to be sufficiently strong to constrain transport coefficients of the hydrodynamic theory.

To implement the second law \eqref{E:local2nd} and \eqref{E:constitutive2nd} we first note that \eqref{E:firstlaw} and \eqref{E:GibbsDuhem} imply
\begin{align}\label{E:IdealEntropyCurrent}
    \nabla_{\mu}\left(su^{\mu}\right)&=-\frac{1}{T}u_{\nu}\nabla_{\mu}T_{id}^{\mu\nu}-\frac{1}{2}\frac{\mu_{ab}}{T}\nabla_{\mu}S_{id}^{\mu ab}-\frac{\mu}{T}\nabla_{\mu}J_{id}^{\mu}\,.
\end{align}
Now consider $T_{C,id}^{\mu\nu} = T_{id}^{\mu\nu} - B^{\mu\nu}[S_{id}]$. Using
\begin{equation}
	\frac{1}{2}\frac{u_{\nu}}{T}\nabla_{\mu}\nabla_{\lambda}\left(S_{id}^{\lambda\mu\nu}-S_{id}^{\mu\lambda\nu}-S_{id}^{\nu\lambda\mu}\right)= \frac{1}{2T}R^{\rho\sigma\nu\lambda}u_{\nu}u_{\rho}\rho_{\lambda\sigma}=0
	\label{E:EntropyIdealContributionS}
\end{equation}
allows us to replace $T_{id}^{\mu\nu}$ in \eqref{E:IdealEntropyCurrent} with $T_{C,id}^{\mu\nu}$. Using 
\begin{equation}
\label{E:Roi2}
	\mu_{\alpha\beta} T_{id}^{[\alpha\beta]} = 0
\end{equation}
and $B^{[\mu\nu]}[S] = \frac{1}{2} \nabla_{\lambda}S^{\lambda\mu\nu}$ we find that 
\begin{equation}
	 \nabla_{\mu}\left(su^{\mu}\right)=-\frac{1}{T}u_{\nu}\nabla_{\mu}T_{C,id}^{\mu\nu} + \frac{\mu_{\alpha\beta}}{T}T_{C\,id}^{[\alpha\beta]}-\frac{\mu}{T}\nabla_{\mu}J_{id}^{\mu}\,.
\end{equation}
Therefore, if we define the canonical entropy current
\begin{equation}
	J_c^{\mu} = s u^{\mu} - \frac{u_{\nu}}{T} T_{C,d}^{\mu\nu} - \frac{\mu}{T} j_{d}^{\mu}
\end{equation}
with
\begin{align}
\begin{split}
	T_{C,d}^{\mu\nu} &= T_C^{\mu\nu}-T_{C,id}^{\mu\nu} - T_A^{\mu\nu} \\
		&= -\eta \sigma^{\mu\nu} - \zeta \nabla_{\alpha}u^{\alpha} P^{\mu\nu} + \mathcal{O}(\nabla^2)\,,
	\\
	j_d^{\mu} &= J^{\mu} - J_{id}^{\mu} - J_A^{\mu} \\
		&= \sigma_E \mathcal{E}^{\mu} + \mathcal{O}(\nabla^2)\,,
\end{split}
\end{align}
(under the equations of motion) then
\begin{align}
\begin{split}
	\nabla_{\mu}J_c^{\mu} &= - T_{C,d}^{\mu\nu} \left(\nabla_{\mu} \frac{u_{\nu}}{T} + \frac{\mu_{\mu\nu}}{T}\right) + \frac{1}{T} j_d^{\mu} \mathcal{E}_{\mu} \\
	&= \frac{\eta}{2T} \sigma_{\mu\nu}\sigma^{\mu\nu} + \frac{\zeta}{T} \left(\nabla_{\mu}u^{\mu}\right)^2 + \frac{\sigma_E}{T} \mathcal{E}^{\mu}\mathcal{E}_{\mu} + \mathcal{O}(\nabla^3) \,.
\end{split}
\end{align}

The most general entropy current is given by 
\begin{equation}
	J_s^{\mu} = J_c^{\mu} + J_{nc}^{\mu}\,,
\end{equation}
where, at first order in derivatives, $J_{nc}^{\mu}$ is a linear combination of all first order vectors. Repeating the standard entropy analysis of \cite{BhattacharyaBhattacharyyaMinwallaYarom:2011} for a general torsionless background, we find that $J_{nc}^{\mu}=0$. 

Allowing a background contorsion cannot modify this conclusion:
torsionless backgrounds form a subset of the allowed torsionful
backgrounds. A torsionful analysis may require additional contributions
to the entropy current that explicitly contain the contorsion, but all
such terms vanish when the contorsion is set to zero, see \cite{Gallegos:2022jow} for an explicit example with non-vanishing torsion. Hence the entropy
current relevant to the torsionless constitutive relations considered
here is $J_s^\mu=J_c^\mu$.
Thus, for entropy to be positive semi-definite in a torsionless background it must be the case that
\begin{equation}
	\eta \geq 0,\qquad \zeta \geq 0, \qquad \sigma_E \geq 0\,.
\end{equation}

\section{Decomposing contorsion}\label{A:contorsion}
The contorsion tensor has mixed symmetry. To simplify the expressions associated with it, it is convenient to decompose it into Lorentz representations, viz., 
\begin{equation}
    K_{cab} =\mathcal{K}_{cab}+K_{[cab]}+\frac{2}{3}\eta_{c[a}K_{b]}
\end{equation}
where
$\mathcal{K}_{[cab]}= 0$ and $\tdu{\mathcal{K}}a{ab}=0$. 
and then further decompose each of these relative to the $SO(3)$ residual symmetry of the directions orthogonal to $u^{\mu}$, 
\begin{equation}
\begin{aligned}
	k_{A}^{ab}&=u_{c}K^{[abc]}
	&\tilde{\kappa}&=\frac{1}{6}\eps_{abcd}u^{d}K^{[abc]}
	\\
	k_{V}^{a}&=\tud PabK^{b}
	&\kappa&=u_{b}K^{b}
	\\
	\mathcal{K}_{A}^{ab}&=P^{ac}P^{bd}u^{e}\mathcal{K}_{ecd}
	&\mathcal{K}_{S}^{ab}&=\frac{1}{2}P^{bd}u_{c}\tdu{\mathcal{K}}d{ac}+\left(a\leftrightarrow b\right)
	\\
	\mathcal{K}_{V}^{b}&=u^{a}u^{c}P^{bd}\mathcal{K}_{acd}
	&\mathcal{K}_{Tabc}&=\tdu Pad\tdu Pbe\tdu Pcf\mathcal{K}_{def}\,.
\end{aligned}
\end{equation}
Note that this decomposition is different from the one used in \cite{Gallegos:2022jow},\footnote{In \cite{Gallegos:2022jow} $\kappa_V{}^\mu$ and $\kappa_T{}^{\alpha \beta \mu}$ were denoted by $\mathcal{K}_V{}^\mu$ and $\mathcal{K}_T{}^{\alpha \beta \mu}$ respectively. To avoid confusion with the decomposition used in this work, their names were adjusted to what is shown in \eqref{contDecompOriginal}. All other components kept the names introduced in \cite{Gallegos:2022jow}.} where torsion was decomposed as 
\begin{equation}
    \begin{split}\label{contDecompOriginal}
       K_{cab} &= - u_c (u^a k^b - u^b k^a + K^{ab}) + \frac{2 \kappa}{3} u^{[a} \Delta^{b]}{}_c+ 2 u^{[a} \kappa_S{}^{b]}{}_c + 2 u^{[a} \kappa_A{}^{b]}{}_c \, \\  &+ \frac{1}{2} \Delta_c{}^{[a} \kappa_V^{b]}  + \kappa_{T}{}_c{}^{ab} + \mathcal{K}_{Ac}{}^{ab}
    \end{split}
\end{equation}
such that
\begin{equation}
    \begin{aligned}
        k^a &= u^\mu u_b K_\mu{}^{ab} \, , \qquad &K^{ab} &= \Delta^a{}_c \Delta^b{}_d u^\mu K_\mu{}^{cd} \, , \\
        \kappa &= K_c{}^{cd}u_d \, , \qquad &\kappa_S{}^{\mu \nu} &= \left( \Delta^{(\mu}{}_c \Delta^{\nu)\rho}- \frac{1}{3}\Delta^{\mu \nu}\Delta^\rho{}_c \right) K_\rho{}^{cd} u_d \, , \\ \kappa_A^{\mu \nu} &= \Delta^{[\mu}{}_c\Delta^{\nu]\rho} K_\rho^{cd}u_d \, , \qquad &\kappa_V^\mu &= 2 \Delta^\mu{}_d \Delta^\rho{}_c K_\rho{}^{cd} \, , \\ 
        \mathcal{K}_A{}^{\lambda \rho \sigma} &= \Delta^{[\rho}{}_c \Delta^\sigma{}_d \Delta^{\lambda] \alpha}K_\alpha{}^{cd} \, , \qquad & \kappa_{T \mu}{}^{\alpha \beta} &= \Delta^\alpha{}_c \Delta^d{}_\beta \Delta^\nu{}_\mu K_\nu{}^{cd} - \frac{1}{2} \Delta^{[\alpha}_\mu \kappa_V^{\beta]} - \mathcal{K}_{A \mu}{}^{\alpha \beta}
    \end{aligned}
\end{equation}
The two decompositions are related through the transformations
\begin{equation}
\begin{aligned}
k_{A}^{ab}&=\frac{1}{3} K^{ab} - \frac{2}{3} \kappa_A{}^{ab} \, , &\tilde{\kappa}&= \frac{1}{6} \epsilon_{\beta \mu \nu\alpha} \mathcal{K}_{A}{}^{\beta \mu \nu }  u^\alpha\, ,\\ 
k_{V}^{a}&=k^a + \frac{1}{2} \kappa_V^a  \, ,    &\kappa&=\kappa \, , \\
\mathcal{K}_{A}^{ab}&= \frac{2}{3} K^{ab} + \frac{2}{3} \kappa_A{}^{ab} \, ,    &\mathcal{K}_{S}^{ab}&=\kappa_S{}^{ab}\, , \\
\mathcal{K}_{V}^{b}&=- \frac{2}{3} k^b + \frac{1}{6} \kappa_V{}^b \,, &\mathcal{K}_{T abc} &= \kappa_{T abc} - \frac{2}{3} \Delta_{a[b}k_{c]}  + \frac{1}{6} \Delta_{a [b} \kappa_{V c]}
\end{aligned}
\end{equation}

\section{Computing $\rho_M$}
\label{A:rhoMtable}
This appendix records both the raw results used in table \ref{T:various} and the conversion to the conventions of this paper. 

\subsection{Free fields}
\label{AA:freefields}
According to \cite{KovtunShukla:2018} the hydrostatic partition function of a conformal, parity preserving theory is given by
\begin{equation}
 W=\int d^4x\sqrt{-g}\left[P_0+f_1\left( R+6 a_{\alpha}a^{\alpha} \right) -6 \frac{\partial f_1}{\partial{\mu} }E_{\alpha}a^{\alpha} +f_3\tilde{\Omega}_\mu\tilde{\Omega}^\mu+\ldots \right]\,.
 \label{E:appendixKSfunctional}
\end{equation}
(See (2.9) and (2.12) of \cite{KovtunShukla:2018}).
Comparing \eqref{E:appendixKSfunctional} to \eqref{E:WnoK} and using
\begin{equation}
	\Omega^{\mu\alpha}\Omega^{\nu}{}_{\alpha} = \frac{1}{4}\left( P^{\mu\nu} \tilde{\Omega}^2 - \tilde{\Omega}^{\mu} \tilde{\Omega}^{\nu} \right) 
\end{equation}
we find that
\begin{equation}
	 \kappa_g=-2f_1,
	 \qquad
	\lambda_3^h = 2 f_3
\end{equation}
implying
\begin{equation}
	\lambda_3 = -8 f_3 + 4 f_1  + \frac{\left(4 c_A \mu^3 + 32 \pi^2 c_m T^2 \mu \right)^2}{4 P_0}\,.
\end{equation}

For one real (uncharged) scalar, Kovtun and Shukla obtain (see equation (3.2) of \cite{KovtunShukla:2018})
\begin{equation}
 f_1=\frac{T^2}{144}(1-6\xi),
 \qquad
 f_3=-\frac{T^2}{144},
 \label{E:appendixScalarRaw}
\end{equation}
where $\xi$ is the curvature coupling of the scalar field to the Ricci scalar. In a conformal theory we set  $\xi=1/6$ and obtain
\begin{equation}
 \kappa_g=0,\qquad \lambda_3=\frac{T^2}{18},\qquad
 \rho_M=\frac{T^2}{72}\,.
\end{equation}
For a four-component massless Dirac fermion at zero chemical potential, the  result is
\begin{equation}
 f_1=-\frac{T^2}{144},
 \qquad
 f_3=-\frac{T^2}{288}.
 \label{E:appendixDiracRaw}
\end{equation}
(see (3.6) of \cite{KovtunShukla:2018}).
Thus 
\begin{equation}
	\kappa_g = \frac{T^2}{72},\qquad
	\lambda_3 =0,\qquad
	\rho_M = \frac{T^2}{144}
\end{equation}
For one Maxwell field, the corresponding susceptibilities are
\begin{equation}
	 f_1=-\frac{T^2}{36},
	 \qquad
	 f_3=\frac{T^2}{36}.
 \label{E:appendixMaxwellRaw}
\end{equation}
(See (3.8) of \cite{KovtunShukla:2018}).)
Consequently
\begin{equation}
	 \kappa_g=\frac{T^2}{18},\qquad
	 \lambda_3=-\frac{T^2}{3},\qquad
	 \rho_M=-\frac{T^2}{18}.
\end{equation}
Note that since we are dealing with free fields the expressions for transport is additive.

For a massive free Dirac field at $T=0$ and vector chemical potential, $|\mu_V|>m$, one finds via equations (42) and (44) of \cite{Shukla:2019} that
\begin{align}
 f_1&=-\frac{1}{48\pi^2}\left(
 |\mu_V|\sqrt{\mu_V^2-m^2}-m^2\log\left(\frac{|\mu_V|+\sqrt{\mu_V^2-m^2}}{m}\right)\right),
 \nonumber\\
 f_3&=-\frac{|\mu_V|\sqrt{\mu_V^2-m^2}}{96\pi^2}.
 \label{E:appendixDenseDiracRaw}
\end{align}
And $f_1=f_3=0$ for $|\mu_V|<m$.  The relation between $\rho_M$ and the conformal coefficients should be imposed only after taking $m\to0$.  In that limit,
\begin{equation}
 f_1=-\frac{\mu_V^2}{48\pi^2},\qquad
 f_3=-\frac{\mu_V^2}{96\pi^2},\qquad
 \kappa_g=\frac{\mu_V^2}{24\pi^2},\qquad
 \lambda_3=0,\qquad
 \rho_M=\frac{\mu_V^2}{48\pi^2}.
\end{equation}
Since $\mu_V$ is a vector chemical potential $c_A=c_m=0$.

\subsection{Holography}

For the neutral plasma dual to two-derivative Einstein the authors of \cite{BaierRomatschkeSonStarinetsStephanov2008,BhattacharyyaHubenyMinwallaRangamani2008}
find
\begin{equation}
\label{E:gotkappag}
 \kappa_g=\frac{\eta}{\pi T},
 \qquad
 \lambda_3=0.
\end{equation}
For planar $\mathcal N=4$ SYM at infinite 't Hooft coupling, $\eta=\pi N_c^2T^3/8$, and hence
\begin{equation}
 \kappa_g=\frac{N_c^2T^2}{8},
 \qquad
 \rho_M=\frac{N_c^2T^2}{16}.
\end{equation}
This row is neutral, so anomaly coefficients are not required. 

For the plasma dual to the Einstein-Maxwell-Chern-Simons theory, 
\begin{align}
	S = -\frac{1}{16\pi G_5} \int \sqrt{-G} \left(R+12-\frac{1}{4} F_{MN}F^{MN}\right)d^5x + S_{CS}\,,
\intertext{with}
	S_{CS} = \frac{\mathcal{C}}{16\pi G_5} \int \frac{1}{12 \sqrt{3}} \epsilon^{MNPQR}A_M F_{NP}F_{QR} d^5x\,,
\end{align}
(with $\mathcal{C}=1$ for $\mathcal{N}=4$ super Yang Mills)
the authors of \cite{Erdmenger:2008rm,Banerjee:2008th} obtained
\begin{equation}
	\lambda_3
	 =\frac{N_c^2 \mathcal{C}^2 \mu^2 b^4 r_+^4}{6\pi^2}
	 \label{E:appendixEHKYraw}
\end{equation}
where
\begin{equation}
	r_+ = \frac{\pi T}{2}\left(1+q \right)\,,
	\qquad
	b^{-4} =  \frac{\pi^4 T^4}{2^4} \left(q+1\right)^3 (3q-1)
\end{equation}
with
\begin{equation}
	q= \sqrt{1+\frac{2\mu^2}{3\pi^2T^2}}\,.
\end{equation}
(Note that in the conventions of \cite{Erdmenger:2008rm} $\Sigma^{(3))}_{\mu\nu}=-\Omega^{\alpha\langle\mu}\Omega_{\alpha}{}^{\nu\rangle}$.) The expression for $\kappa_g$ was computed in \cite{GrieningerShukla:2021} using the same notation as in section \ref{AA:freefields}. There (equation (3.5) of \cite{GrieningerShukla:2021}) the authors compute
\begin{equation}
	f_1 = -\frac{N_c^2}{8\pi^2} \frac{r_+^2}{2}\,.
\end{equation}
(where we have restored the overall normalization of $N_c^2/(8\pi^2)$)
or
\begin{equation}
	\kappa_g = \frac{N_c^2 r_+^2}{8\pi^2}  
\end{equation}
(We have inferred the overall normalization by taking the $\mu\to 0$ limit of $\kappa_g$ and comparing it to $-\kappa_g/2$ in \eqref{E:gotkappag}.) 

Finally, the anomaly $c_A$ is associated with the Chern Simons term in the bulk action. In the conventions of \cite{Erdmenger:2008rm},
The on shell action is the generating function for connected correlators, so the Ward identity for the boundary consistent current, $J_{cons}^{\mu}$, can be obtained via
\begin{equation}
	\delta_{\alpha}S = \delta_{\alpha}S_{CS} = \int \sqrt{-g} J_{cons}^{\mu} \partial_{\mu} \alpha d^4x = -\int \sqrt{-g} \alpha \nabla_{\mu}J_{cons}^{\mu} d^4x \,.
\end{equation}
Using $G_5 = \frac{\pi}{2 N_c^2}$ (in units where the AdS radius is one), we find
\begin{equation}
\label{E:consanomaly}
	\nabla_{\mu}J_{cons}^{\mu} = \frac{\mathcal{C}N_c^2}{96 \sqrt{3}\pi^2} \epsilon^{\mu\nu\rho\sigma}F_{\mu\nu}F_{\rho\sigma}\,.
\end{equation}
up to an overall sign depending on the oritentation convention of the $\epsilon$ tensor (which will not be relevant for us since we will eventually only need $c_A^2$).
The covariant current, $J^{\mu}$ is obtained by adding to the consistent current a Bardeen Zumino term, 
\begin{equation}
	J^{\mu} = J_{cons}^{\mu} + c_A \epsilon^{\alpha\beta\mu\nu}F_{\alpha\beta}A_{\nu}\,,
\end{equation}
see \cite{Jensen:2012kj}. Comparing the divergence of the consistent anomaly \eqref{E:consanomaly} with the covariant one \eqref{E:covariantanom}, we find  
\begin{equation}
	c_A = \frac{\mathcal{C}N_c^2}{24 \sqrt{3} \pi^2}\,.
\end{equation}
From a holographic perspective, the mixed anomaly has been set to zero. The mixed gauge-gravitational anomaly is absent in this holographic model so we set $c_m=0$. Since adding a mixed gauge--gravitational Chern--Simons term would modify second-order transport coefficients such as $\lambda_3$ \cite{Megias-Pena:2013}; we will not add such a term by hand.

The last ingredient we need in order to evaluate the spin density is the pressure. From \cite{Erdmenger:2008rm} we have
\begin{equation}
	P_0 = \frac{N_c^2}{8\pi^2 b^4}\,.
\end{equation}
Putting all this together we find
\begin{equation}
	\rho_M = \frac{N_c^2 T^2}{64} \left((1+q)^2 -4 \mathcal{C}^2(q-3)(q-1)\right)  \,.
\end{equation}
Setting $\mathcal{C}=1$ for $\mathcal{N}=4$ super Yang Mills we obtain the expression in table \ref{T:various}. (If we want to describe a non Yang-Mills theory we may keep $\mathcal{C}\neq 1$ and treat $N_c^2$ as the number of colors up to an overall numerical factor associated with the volume of the compact manifold relative the $S^5$. )

We note in passing that, in the conventions of \cite{GrieningerShukla:2021} where $N_c^2/(8\pi^2)=1$, we have
\begin{equation}
	f_3 = -\frac{\rho_M}{2}\,.
\end{equation}
Expanding $f_3$ in powers of $\mu/T$ we find
\begin{equation}
 \frac{f_3}{T^2}=-\frac{\pi^2}{4}-\frac14\left(\frac{\mu}{T}\right)^2
 +O\left(\frac{\mu^4}{T^4}\right).
 \label{E:appendixGScheck}
\end{equation}
The leading term in the expansion is consistent with the $\mu=0$ solution we discussed earlier. The subleading term in the expansion
can be compared to the subleading term in the numerical expression for $f_3$ obtained in \cite{GrieningerShukla:2021}, $-\pi^2/4-0.2489(\mu/T)^2$, in good agreement with the analytic coefficient.

The authors of \cite{GrozdanovStarinets:2017} considered the theory dual to 
\begin{equation}
	S = \frac{1}{2\kappa_5^2} \int \sqrt{-G} \left(R+12+\frac{\lambda_{GB}}{2}\left(R^2 - 4 R^{MN}R_{MN}+R_{MNPQ}R^{MNPQ}\right) \right)d^5x
\end{equation}
Using equations (4.20) and (4.23) of \cite{GrozdanovStarinets:2017}, together with \eqref{E:rhoMlambda3}, one has
\begin{align}
\begin{split}
\label{E:GBfull}
	\kappa_g &= \frac{\pi^2  T^2}{\sqrt{2}\kappa_5^2}  \frac{2\gamma_{GB}^2-1}{\sqrt{1+\gamma_{GB}}}  \,, \\
	\lambda_3 &= -\frac{2\pi^2  T^2}{\sqrt{2}\kappa_5^2} \frac{3+\gamma_{GB}-4 \gamma_{GB}^2}{\sqrt{1+\gamma_{GB}}}\,, \\
	\rho_M &= \frac{\pi^2 T^2}{2 \sqrt{2}\kappa_5^2} \frac{6 \gamma_{GB}^2 -4-\gamma_{GB}}{\sqrt{1+\gamma_{GB}}}\,,
\end{split}
\end{align}
with
\begin{equation}
	 \gamma_{GB}=\sqrt{1-4\lambda_{GB}}\,.
\end{equation}
Early holographic Weyl-anomaly computations relating curvature-squared couplings to $a$ and $c$ were carried out in \cite{Nojiri:1999mh,Blau:1999vz}. In the present Gauss--Bonnet normalization, equations (1.1) and (1.4) of \cite{Kats:2007mq} give,
\begin{equation}
	\lambda_{\rm GB}=\frac{c-a}{4c}
	+\mathcal O \left(\left(\frac{c-a}{c}\right)^2\right)\,.
\end{equation}
For $\mathcal{N}=4$ super Yang Mills, $a=c$ and therefore the Gauss-Bonnet correction vanishes to leading order. 

If we use instead the $\mathcal{N}=2$ $Sp(N_c)$ theory described in \cite{Blau:1999vz,Kats:2007mq}, then equations (2.10) of \cite{Blau:1999vz} read
\begin{equation}
	a = \frac{12 N_c^2 + 12 N_c -1}{24} \,,
	\qquad
	c= \frac{6 N_c^2+9 N_c - 1}{12}\,,
\end{equation}
and therefore, in this setup
\begin{equation}
	\lambda_{GB}=\frac{1}{8N_c}+\mathcal O(N_c^{-2})\,.
\end{equation}
To relate the gravitational coupling $\kappa_5^2$ to the number of colors $N_c$ we use equations (2.1), (2.6), (2.9) and (2.10) of \cite{Myers:2010tj} which imply
\begin{equation}
	\kappa_5^2 = \frac{\pi^2 \gamma_{GB} (1+\gamma_{GB})^{3/2}}{2 \sqrt{2} c}
	= \frac{2\pi^2}{N_c^2} \left( 1-\frac{31}{16 N_c} + \mathcal{O}(N_c^{-2})\right)
\end{equation}
Expanding \eqref{E:GBfull} perturbatively in $1/N_c$ gives the $Sp(N_c)$ result quoted in table \ref{T:various}.

\end{appendix}

\bibliographystyle{JHEP}
\bibliography{hydrospinbib}

\end{document}